\documentclass[iop, apj, tighten, onecolumn]{emulateapj}
\usepackage[]{amsmath}
\usepackage{float}

\newcommand{\mufmr}{$\mu_{\mathrm{FMR}} = \mathrm{Log(mass)} - \alpha \times \mathrm{Log(SFR)}$ }

\begin{document}

\submitted{ ApJ Referee response submitted December 6, 2012 }
\title{The Fundamental Metallicity Relation Reduces Type Ia SN Hubble Residuals More Than Host Mass Alone}
\author{Brian T. Hayden\altaffilmark{1},
Ravi R. Gupta\altaffilmark{2},
Peter M. Garnavich\altaffilmark{1},
Filippo Mannucci\altaffilmark{3},
Robert C. Nichol\altaffilmark{4},
Masao Sako\altaffilmark{2},
}
\altaffiltext{1}{University of Notre Dame, 225 Nieuwland Science Hall, Notre Dame, IN 46556, USA}
\altaffiltext{2}{Department of Physics and Astronomy, University of Pennsylvania, 209 South 33rd Street, Philadelphia, PA 19104, USA}
\altaffiltext{3}{Istituto Nazionale di AstroÞsica, Osservatorio AstroÞsico di Arcetri Largo E Fermi 5 50125 Firenze, Italy}
\altaffiltext{4}{Institute of Cosmology and Gravitation, Portsmouth University, Dennis Sciama Building, Po1 3FX, Portsmouth, UK}

\begin{abstract}
Type Ia Supernova Hubble residuals have been shown to correlate
with host galaxy mass, imposing a major obstacle for their use
in measuring dark energy properties. Here, we calibrate the
fundamental metallicity relation (FMR) of \citet{man10} for
host mass and star formation rates measured from broad-band colors
alone. We apply the FMR to the large number of hosts from the SDSS-II
sample of \citet{gup11} and find that the scatter in the Hubble residuals is
significantly reduced when compared with using only stellar mass (or the mass-metallicity relation) as a fit
parameter.  Our calibration of the FMR is
restricted to only star-forming galaxies and
in the Hubble residual calculation we include
only hosts with log(SFR) $> -2$. Our results strongly suggest that metallicity is the underlying source of the correlation between Hubble residuals and host galaxy mass. Since
the FMR is nearly constant between z = 2 and the present, use of the
FMR along with light curve width and color should provide
a robust distance measurement method that minimizes systematic
errors.
\end{abstract}


\section{Introduction}
\label{sec:intro}
Type Ia supernovae (SNe Ia) are important for their use as distance indicators. They are not perfect standard candles, however; the width of the light curve correlates with the peak brightness (i.e. the Phillips relation, \citealt{phil93}), and color can be used to correct for dust extinction \citep{rie96}. These correlations are instrumental in the use of SNe Ia to measure the cosmological properties of our universe, and these measurements have shown that our universe will not only expand forever \citep{garn98a, perl98}, but that the expansion is actually accelerating \citep{rie98, perl99, knop03, ton03, bar04, rie04}. This acceleration, driven by a ``dark energy" (so called because the source is unknown), has been one of the most surprising scientific discoveries of recent history. 

Cosmological studies have shifted from confirming acceleration to constraining the properties of dark energy \citep{garn98b, ast06, rie07, wood07, mik07, eisen07, con11, sul11, suz12}. Yet as the number of SNe Ia has increased, the statistical error in the cosmological parameters has been decreased such that systematic errors have begun to dominate \citep{wood07, kess09, guy10}. The focus of SN Ia science has thereby shifted to constraining other parameters that may affect their peak brightness. One such parameter that appears to be correlated with Hubble residuals is the stellar mass of the host galaxy \citep{kel10, lam10, sul10, gup11}. \citet{sul06} also showed that the SN Ia rate was dependent on the host galaxy stellar mass and star formation rate (SFR), indicating an environmental effect. While the underlying source of this correlation is unknown, logic suggests that the stellar mass must be tracing some underlying property, such as the progenitor metallicity \citep{tim03, kas09} or age \citep{kru10}. 

Theory does predict that the amount of Ni-56 produced may depend on the metallicity of the progenitor \citep{tim03, kas09}, in the sense that more metal-rich environments produce a higher fraction of stable Ni-58 (due to excess available neutrons) and therefore less luminous SNe Ia. Since the light curve is driven by the decay of Ni-56, an increase in the amount of stable Ni-58 produced will decrease the amount of Ni-56, and should reduce the peak brightness of the SN and change the shape of the light curve. Most of this correlation should be accounted for by the Phillips relation, but with some residual error due to small predicted variations in the Phillips relation at different metallicities \citep{kas09}. \citet{kas09} predicts that the systematic error on distance measurements due to this effect should be around 2\%. \citet{dandrea11} measured gas-phase metallicities of SN Ia host galaxies directly and found a strong correlation (4$\sigma$) with Hubble residuals\footnote{Hubble residuals are the residual scatter in the SN distance modulus after taking into account sources of peak brightness variation}, suggesting that the stellar mass correlation could be a tracer of the metallicity. \citet{gal08} studied absorption line strengths in early-type host galaxies and found both an age and a metallicity correlation with Hubble residuals. 

Recent work by \citet{man10}, using Sloan Digital Sky Survey (SDSS) DR7 catalogs containing emission line fluxes and other spectroscopic properties of galaxies\footnote{http://www.mpa-garching.mpg.de/SDSS/}, discovered that galaxy metallicity is correlated not only to stellar mass, but also to SFR. Using the SFR in addition to the mass as a basis to estimate a galaxy's metallicity can significantly increase the precision of the metallicity estimate compared to mass alone (i.e. the Tremonti relation, \citealt{tre04}). \citet{man10} found that star-forming galaxies up to z=2.5 follow the same local relation between metallicity, mass, and SFR, with no evidence for evolution. Such a correlation, termed the fundamental metallicity relation (FMR) by \citet{man10}, can improve metallicity estimates when direct measurement of metallicity from relevant emission lines is not possible. While the FMR was originally defined for stellar masses above $10^{9.5}$ M$_{\odot}$, it was later extended towards lower masses \citep{man11}. Among others, \citet{yates12} and \citet{per12} obtained qualitatively very similar relations between metallicity, mass and SFR, although with some quantitative differences. 

For stellar masses above about $10^{10.5}$ M$_{\odot}$, the metallicity is nearly constant. Recent work by \citet{joh12} found very little Hubble residual correlation with stellar mass for host galaxies with mass $> 10^{10.5}$ M$_{\odot}$. This lack of correlation at higher masses could indicate that the mass is tracing the metallicity, considering the lack of variation with Hubble residuals in this region and the constancy of the metallicity.  A significant goal of this paper is to test the ability of stellar population synthesis models to reproduce the FMR, which would then allow for accurate metallicity estimates using multi-band photometry alone. 

In this paper, we use the MPA/JHU catalogs of galaxy properties to calibrate the fundamental metallicity relation for use with two multi-band photometric stellar population synthesis fitters. We present our method and results using Z-PEG \citep{zpeg}, as well as the fitting method of  \citet{gup11} using the Flexible Stellar Population Synthesis (FSPS) models \citep{fsps, fsps2}. Hereafter, we will refer to the fits using the \citet{gup11} fitter as G11. The use of the FMR based on photometry alone allows accurate metallicity estimates of faint galaxies, or galaxies that are otherwise difficult to measure spectroscopically (for instance, high redshift galaxies where the relevant emission lines are redshifted beyond the capabilities of ground-based spectroscopy). This also allows for a study of the SDSS-II SNe Ia using the FMR, as the large sample of several hundred SNe Ia contains only about 40 galaxies that have emission-line estimated metallicities \citep{dandrea11}. Using the improved metallicity estimates based on these photometric measurements, we test the Hubble residuals of the SDSS-II SNe Ia for their correlation with host metallicity. 

In section \ref{sec:data}, we discuss the selection of the SDSS-II host galaxies, and the selection of the SDSS-II star-forming galaxies from the MPA/JHU catalogs. In section \ref{sec:analysis}, we discuss the calibration of the FMR using Z-PEG and G11, and report our results of testing the Hubble residuals against the metallicity parameter from the FMR. In section \ref{sec:discussion} we discuss some issues with the photometric fitters used in the paper, as well as extending the FMR to lower metallicities and higher redshifts. Finally, we present our conclusions in section \ref{sec:conclusions}.

\section{Data}
\label{sec:data}
\subsection{SDSS host galaxy sample}
For this paper we have used the 206 SDSS-II host galaxies from \citet{gup11}. While a detailed summary of the exact method is presented in \citet{gup11}, we will briefly summarize the SDSS-II SN Survey and the selection of the host galaxies in this section.

The SDSS-II SN Survey searched 300 deg$^2$ of sky, known as stripe 82, during 3 month periods of 2005, 2006, and 2007 \citep{frie08} with a cadence of $\approx$ 2 days using the CCD camera on the SDSS 2.5 m telescope \citep{york00, gunn98, gunn06}. Around 500 type Ia supernovae were spectroscopically confirmed over the course of the survey \citep{sako08, hol08}. The SNe Ia range in redshift from 0.01 to 0.42, with a median around 0.2. 

We use the publicly available Supernova Analysis package SNANA \citep{kess09b}, using the light-curve fitting process SALT2 \citep{guy07} to determine the best-fit distance modulus for each supernova from the SDSS-II photometry \citep{hol08, sako11}. Initial cuts on the sample of SNe were made as follows:
\begin{enumerate}
\item at least one measurement with $T_{rest} < -2$ days, where $T_{rest}$ is defined as the number of days from the date of maximum;
\item at least one measurement at $T_{rest} > +10$ days;
\item at least one measurement with signal-to-noise ratio (S/N) $>$ 5 for each of the \textit{g, r} and \textit{i} bands;
\item best-fit probability greater than 0.001, calculated from the $\chi^2$ value of the SALT2 fit.
\end{enumerate}
These cuts reduce the sample to 319 SNe.

Each SN was visually inspected for host galaxy identification. 14 SNe were removed from the sample because they do not have identifiable hosts, either because they fall outside of the SDSS footprint, were too faint, or had \textit{r}-band magnitudes outside the range $15.5 <$ \textit{r} $< 23$. After light-curve fitting and host inspection the sample contains 305 objects. The final cut is the best-fit probability from the \citet{gup11} fits. Fits with probability $< 0.001$ were removed, which leaves 206 SNe in the SDSS-II sample.

\subsection{SDSS general galaxy sample}
\label{catalog}
For comparison with \citet{man10}, we have used the MPA/JHU catalogs for the SDSS-DR7, available at http://www.mpa-garching.mpg.de/SDSS/ and described in \citet{kauf03b}, \citet{brinch04}, and \citet{sal07}. The catalogs contain 927,552 galaxies with measured spectroscopic properties, as well as stellar masses (measured photometrically), multi-band photometry, and other important parameters. We have applied similar cuts to the data in order to reproduce the sample used by \citet{man10} in the calculation of the fundamental metallicity relation (FMR). Our cuts are as follows:
\begin{enumerate}
\item Redshift between 0.07 and 0.30. As described in \citet{man10}, this ensures that the [O${\mathrm{II}}]\lambda$3727 line falls in the useful spectral range, as this line is important in calculating the metallicity, and that the 3-arcsec aperture of the spectroscopic fiber samples a significant fraction of each galaxy. 
\item The signal-to-noise ratio of H$\alpha$ is $>$ 25.  This requirement leads to a more reliable value of the metallicity. In the catalogs it is mentioned that all H$\alpha$ errors can optionally be scaled by 2.473; we do not perform this error modification as it cuts many more galaxies. If this error scaling is correct, this S/N cut on H$\alpha$ would be roughly 10, instead of 25.
\item The Balmer decrement must be greater than 2.5, and the dust extinction, A$_\mathrm{V}$, must be less than 2.5. The extinction was calculated from the Balmer decrement as: 
\begin{equation}
\mathrm{A}_\mathrm{V} = 3.1 \times 2.1575 \times \mathrm{log_{10}} \left ( \frac{F_{H_\alpha} / F_{H_\beta}}{2.86} \right )
\end{equation}
\item AGN were removed using the criteria in \citet{kauf03a}.
\item Beyond the \citet{man10} selection criteria, we also required that 
\begin{equation*}
[\mathrm{O}{\mathrm{II}}]\lambda3726 + [\mathrm{O}{\mathrm{II}}]\lambda3729 + [\mathrm{O}{\mathrm{III}}]\lambda4959 + [\mathrm{O}{\mathrm{III}}]\lambda5007  > 0,
\end{equation*} 
and H$\alpha$ and H$\beta$ $>$ 0. The sum of these oxygen lines is used in the calculation of the metallicity indicator R23. This ensures that no errors occur when we calculate the logarithm of these values in estimating the metallicity, as we had a small percentage of galaxies which did not pass these cuts. 
\end{enumerate}
We calculate the SFR from the H$\alpha$ emission-line flux, corrected for dust extinction estimated from the Balmer decrement. We use the \citet{kenn98} conversion factor between H$\alpha$ luminosity and SFR. While \citet{man10} corrects this value from a \citet{kr01} IMF to a \citet{chab03} IMF, we do not modify the value because we use the \citet{kr01} IMF in our SED fitting process, following previous SN Ia host galaxy analyses.

For the oxygen gas-phase abundance we use a combination of R23 = ([OII]$\lambda$3727 + [OIII]$\lambda$4958,5007)/H$\beta$ and the [NII]$\lambda$6584/H$\alpha$ ratio, as described in \citet{mai08}. We restrict the values such that log([NII]$\lambda$6584/H$\alpha$) $< - 0.35$ and log(R23) $<$ 0.9, and then select only galaxies where these two abundances are within 0.25 dex of each other. These restrictions resolve the ambiguity in the R23 metallicity estimate. 

Our final sample contains 142,227 galaxies, as compared to 141,825 galaxies in the \citet{man10} sample.

\subsection{Photometric fitting using Z-PEG and FSPS Templates}
For photometric fitting of ugriz photometry of galaxies, we use the code Z-PEG\footnote{http://imacdlb.iap.fr:8080/cgi-bin/zpeg/zpeg.pl} \citep{zpeg}, as well as the FSPS model $\chi^2$ minimization technique of \citet{gup11} (G11). We use version 5.23 of Z-PEG, obtained via private communication from Damien Le Borgne\footnote{Information about PEGASE.2, Z-PEG, and contacting the developers can be found at http://www2.iap.fr/pegase/}. Z-PEG utilizes the PEGASE.2 stellar population synthesis code, using 9 different star formation scenarios to build models of galaxy spectral types from elliptical to starburst. Synthetic spectra are produced at galaxy ages of 0 to 20 Gyr. The code performs a $\chi^2$ minimization over all templates convolved with the SDSS filter functions in order to select the best-fit galaxy model, and the parameters of this model such as stellar mass and SFR are considered the best-fit parameters of the galaxy. Errors on these parameters are calculated from the information of the full $\chi^2$ space \citep{zpeg}. 

For our fits, we generally use the default parameters as described in table 1 of \citet{zpeg}. There are two main differences; we leave out the starburst template, and we average the SFR over the past 500 million years. Both of these differences are commonly applied in the literature for using Z-PEG to fit type Ia supernova host galaxies \citep{sul06, lam10, smith10}. We use the other 8 galaxy evolution scenarios described in \citet{zpeg}, with 200 age steps instead of 69, in order to provide better age resolution for older galaxies (the 69 time step scenarios have a time difference of 1 Gyr after ages of about 5 Gyr). For stellar mass fitting, we utilize both redshift constraints; the first constraint is that the galaxy cannot be older than the age of the universe at the given redshift, and the second is that the galaxy must be at least 11 Gyr old at z = 0. The second constraint is a little more relaxed than that listed in \citet{zpeg}, but still has the effect of requiring the best-fit template to contain old stars. For fitting of the SFR, we remove the age constraint at z=0. This allows templates with higher specific star-formation rates, and increases the diversity of the best-fit templates in sSFR. Testing indicated that this method best matched the SFR and sSFR as measured from emission lines.  We also use the same cosmology as \citet{gup11}, with $\Omega_{\Lambda} = 0.735$, $\Omega_\mathrm{m} = 0.274$, and H$_0 = 0.70$. This cosmology is taken from the SDSS-II SN Survey year 1 analysis \citep{kess09}. Lastly, we use the \citet{kr01} initial mass function, following the method of the previously mentioned type Ia supernova host analyses. 

For the G11 fits, the models were built using the same FSPS parameters as listed in table 1 of \citet{gup11}. A $\chi^2$ minimization is performed to select the best-fit model.

\section{Analysis}
\label{sec:analysis}

Our ultimate goal is to study the correlation of SN Ia Hubble residuals with host galaxy stellar masses, using the fundamental metallicity relation, in order to determine the underlying source of this correlation. We will calibrate the FMR using masses and SFRs from two multi-band SED fitting procedures;  we use multiple fitters to show the robustness of the FMR using different stellar population models.  The use of the FMR in our study is necessary because our sample of host galaxies contains a small fraction of spectroscopically measured metallicities. We will calculate the FMR for each of these fitters, and then independently estimate the effect that both the metallicity converted from the FMR, and the FMR parameter, $\mu_{\mathrm{FMR}}$ = log(mass) $-$  $\alpha \times$ log(SFR), have on the Hubble residuals of our sample. 
%Our goal is to use the metallicity measured from the emission lines in the catalogs to calibrate the FMR using the mass and SFR estimates from photometric fitting techniques. The FMR allows accurate metallicity estimates for galaxies where ground-based spectroscopy is difficult or impossible. We will then apply this photometric calibration to host galaxies in the \citet{gup11} sample. We will compare how the metallicity estimate from this calibration and the stellar mass alone reduces the Hubble residuals of these host galaxies.  We must first show that photometrically estimated galaxy properties can be used to calibrate the FMR. This would reduce the difficulty of metallicity estimates on faint galaxies (at higher redshifts, for instance). If the FMR holds for photometrically estimated SFRs, this should be an improvement over the metallicity estimates from mass alone. 

\subsection{Reproducing the Mannucci FMR}

\begin{figure*}
\centering
\begin{tabular}{cc}
\includegraphics*[scale=0.4]{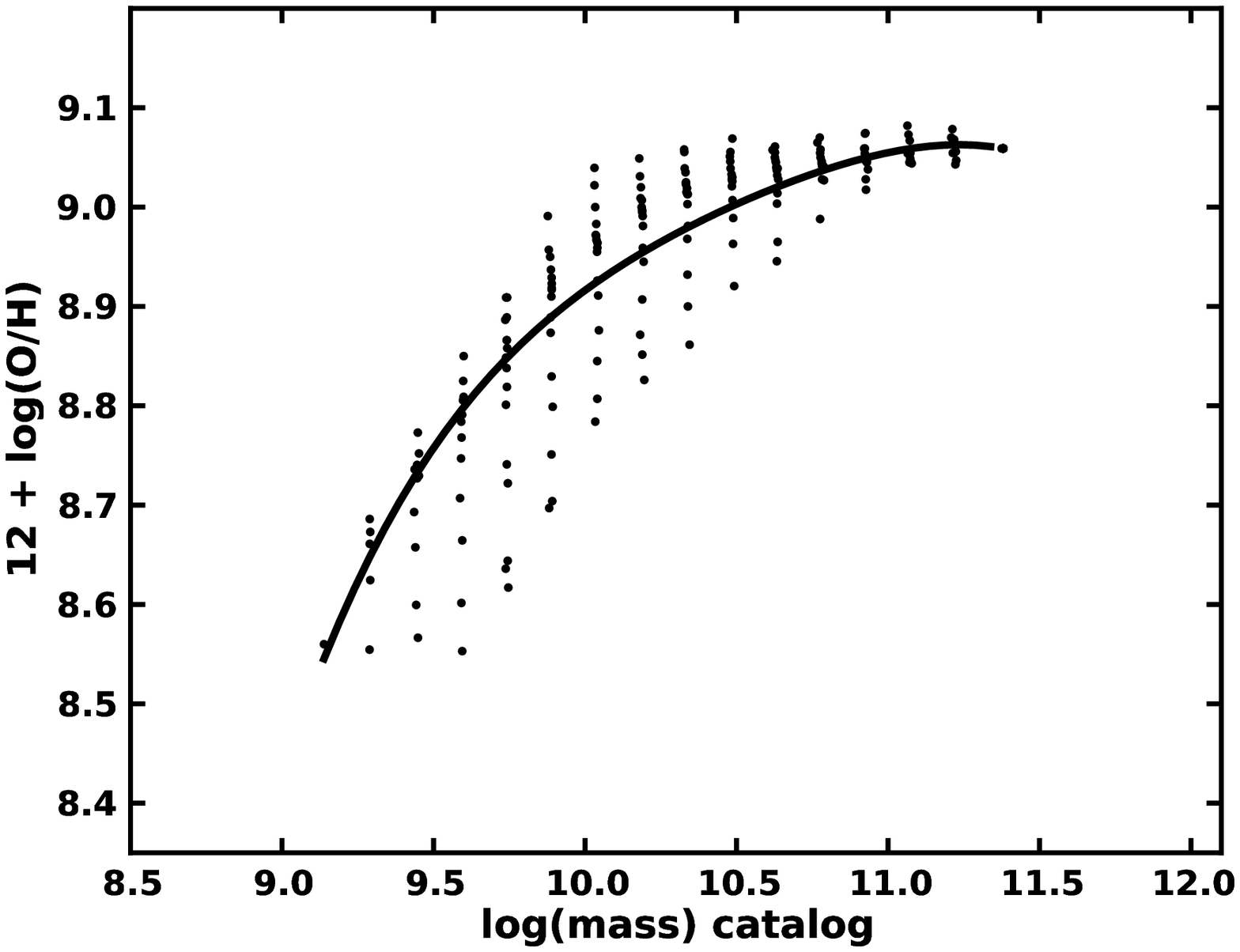} & \includegraphics*[scale=0.4]{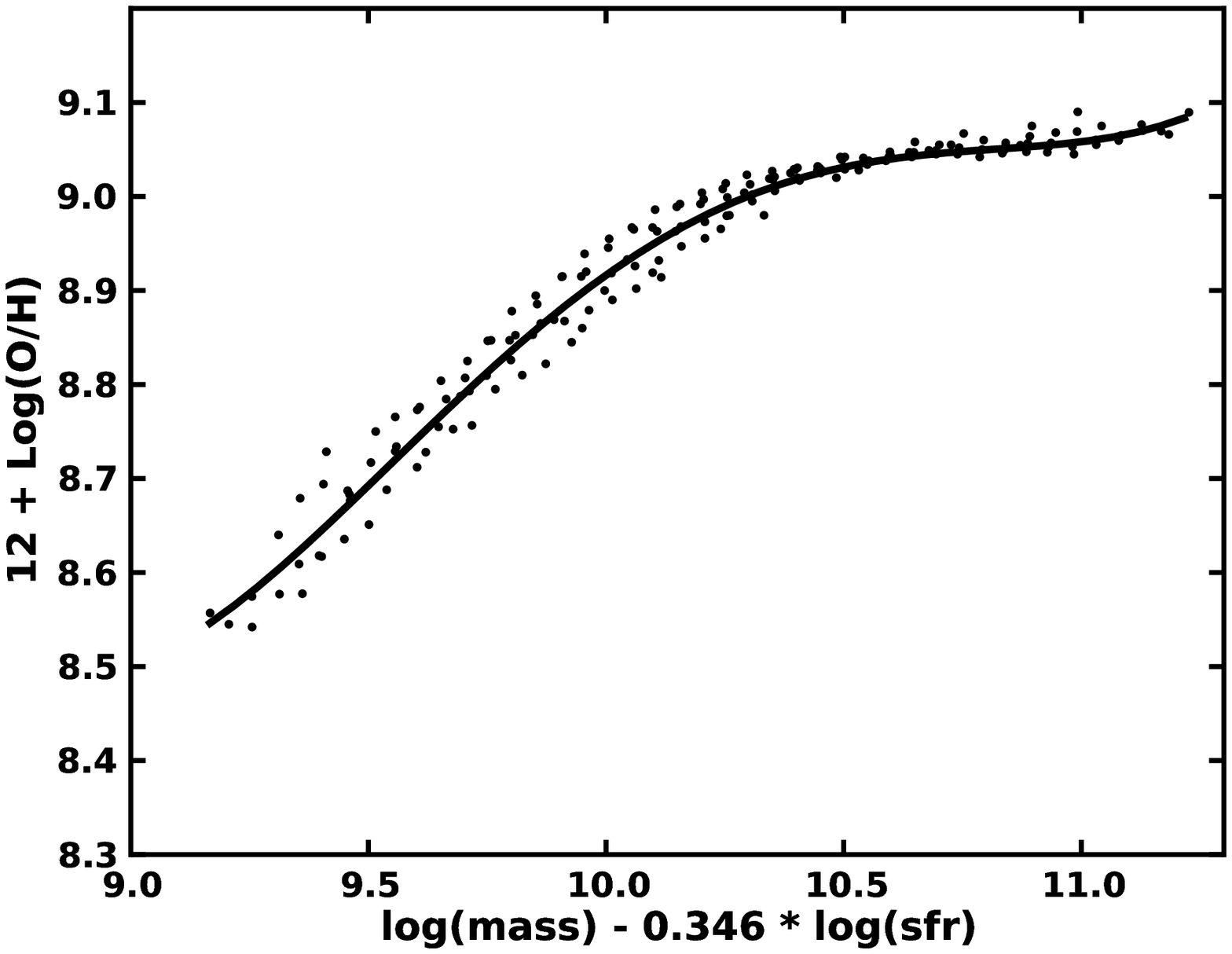}
\end{tabular}
\caption{The fundamental metallicity relation using the stellar masses from the MPA/JHU catalogs. Stellar masses were photometrically estimated, and the SFR is calculated from the H$\alpha$ emission. \textit{Left:} the metallicity of the SDSS-DR7 galaxies, as measured from the MPA/JHU catalog emission lines, as a function of the catalog photometric estimate of the galaxy mass. The points represent the median of the metallicity and the mean of the mass, in bins 0.15 dex wide in the mass.  \textit{Right:} The same objects, also with 0.15 dex wide bins in the emission-line-calculated SFR, showing the reduction in scatter that the SFR provides.  \label{fmr_cat}}
\end{figure*}

\label{man_fmr}
We begin by reproducing the FMR of \citet{man10}. The primary reason for reproducing this analysis is to set the framework and the basic comparison for all photometric formulations of the FMR. For this analysis we use the stellar masses from the MPA/JHU catalogs, and we use the SFR and metallicities as measured from the emission lines in the catalogs, described in section \ref{catalog}.  We bin the 142,227 galaxies in 0.15 dex wide bins in both log(mass) and log(SFR). The metallicity in the bin is estimated by calculating the median value. For the center of each bin in stellar mass and SFR, we use the mean of the values within the bin. This is slightly different from \citet{man10} because we found that the center of the bin was not an accurate representation of the center of the data for bins that were near the edge of the distribution. For the error in metallicity in each bin, we use the standard error of the median, given by $s_e = \frac{1.25 \sigma} {\sqrt{\mathrm{N}}}$, where $\sigma$ is the standard deviation of the metallicity values in the bin and N is the number of galaxies in the bin. Slight differences in building the sample and in the binning process require a re-calibration of the FMR in order to set the basis for comparison with photometric techniques. 
%For the error in metallicity in each bin we use the median absolute deviation, which is given by MAD = $\textrm{median}_i(| n_i - \textrm{median}_j(n_j)|$). This measure of dispersion is less sensitive to a small number of outliers than the standard deviation. Similar to \citet{man10}, we do not use bins that contain less than 50 galaxies.

We then fit a 4th order polynomial to the median metallicity versus mass distribution. This is shown in the left pane of figure \ref{fmr_cat}. This is analogous to the Tremonti relation \citep{tre04}. This 4th order polynomial fit has an RMS dispersion of 0.065 dex. We next fit a 4th order polynomial to the median metallicity versus $\mu_{\mathrm{FMR}}$ distribution, where $\mu_{\mathrm{FMR}} = \mathrm{log(mass)} - \alpha \times \mathrm{log(SFR)}$. We use the label $\mu_{\mathrm{FMR}}$ instead of $\mu$ from \citet{man10}, because in SN Ia cosmology the parameter $\mu$ usually represents the distance modulus. For all FMR and mass-metallicity relation fits reported in this paper, we minimize the $\chi^2$ value of a 4th order polynomial using the Nelder-Mead simplex algorithm minimization in the Python\footnote{http://www.python.org/} module scipy.optimize\footnote{http://docs.scipy.org/doc/scipy/reference/generated/scipy.optimize.fmin.html\#scipy.optimize.fmin}, allowing the FMR parameter $\alpha$ to vary in the fit. The best fit 4th order polynomial and $\alpha$ value are given by:
\begin{equation}
\mathrm{12 + Log(O/H)} = 8.92 + 0.365\mathrm{x} - 0.246\mathrm{x}^2 - 0.081\mathrm{x}^3 + 0.109\mathrm{x}^4\label{eqn_zfmr}
\end{equation}
with x = $\mu_{\mathrm{FMR}}$ $-$ 10, and $\mu_{\mathrm{FMR}}$ = log(mass) $-$ 0.346 $\times$ log(SFR). This best-fit polynomial is shown in the right pane of figure \ref{fmr_cat}, and reduces the RMS dispersion from 0.065 to 0.02 dex. This reduction also appears significant because the metallicity estimate appears to be more gaussian and symmetric about the best-fit line. Our measurement of the FMR matches well with that of \citet{man10}, who found $\alpha = 0.32$ with a minimum dispersion of 0.02 dex.

We have reproduced the results of \citet{man10}, in showing that adding a fraction of the SFR can reduce the scatter in metallicity estimates of galaxies. This formulation of the FMR sets the basis for comparisons with the photometrically estimated FMR. 

%We will now extend this analysis to galaxies with fully photometric estimates of the stellar mass and the SFR, using two different stellar population synthesis models. 

\subsection{Reproducing the FMR with Photometric Fitting Techniques}

\begin{figure*}
\centering
\begin{tabular}{cc}
\includegraphics*[scale=0.4]{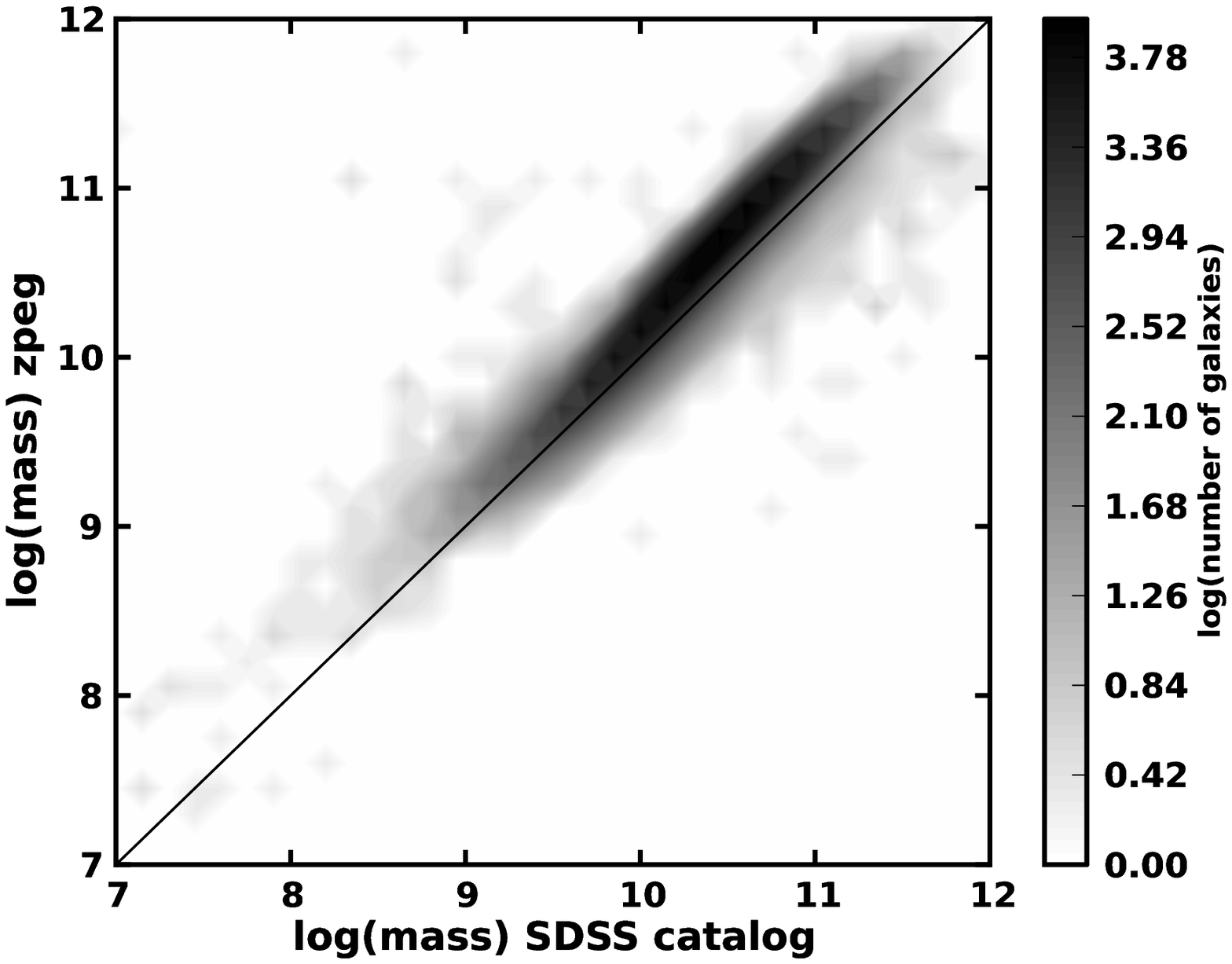} & \includegraphics*[scale=0.4]{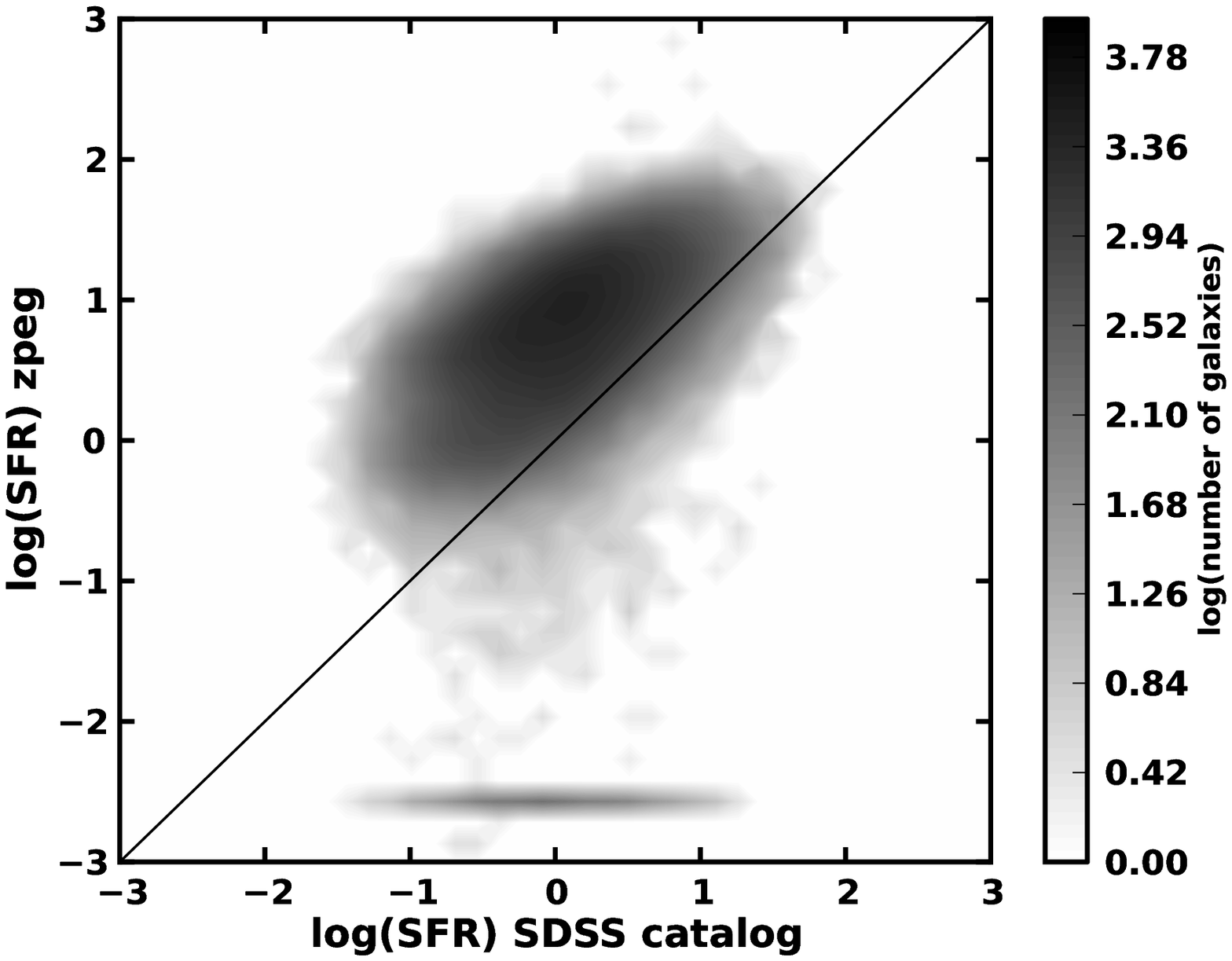}
\end{tabular}
\caption{Density plots (greyscale levels in log space) of the Z-PEG fits compared to the catalog mass and the SFR measured from H$_\alpha$. \textit{Left:} the Z-PEG stellar mass estimates against the SDSS DR-7 mass estimates from \citet{sal07}. The offset is due to the age constraint at z = 0 used in Z-PEG. \textit{Right:} the Z-PEG SFR values against the values measured directly from H$_\alpha$ emission lines. Z-PEG templates with no recent star-formation were assigned an arbitrary value of log(SFR) = $-$2.5 in this figure. These objects are not included in the calculation of the FMR. \label{clouds}}
\end{figure*}

In figure \ref{clouds}, we compare the mass and SFR derived from the Z-PEG fits to the SDSS DR-7 sample. The mass values are highly correlated but have a systematic offset of 0.20 dex as compared to the catalog, which were also calculated photometrically by the method of \citet{sal07}. This offset is caused by the age constraint at z = 0 in the Z-PEG fits. Relaxing the age constraint brings the main group of galaxies closer to the catalog masses, but has the effect of increasing the number of outliers at preferentially low masses. The SFRs as calculated by Z-PEG also show a systematic offset with respect to the SFR measured from H$\alpha$. Z-PEG templates with no recent star formation were assigned an arbitrary value of $-$2.5 for the contour plot. These objects are removed from the sample prior to calculation of the FMR, along with objects with poorly constrained stellar masses, reducing the sample size to 140,987. 

\begin{figure*}
\centering
\begin{tabular}{cc}
\includegraphics*[scale=0.4]{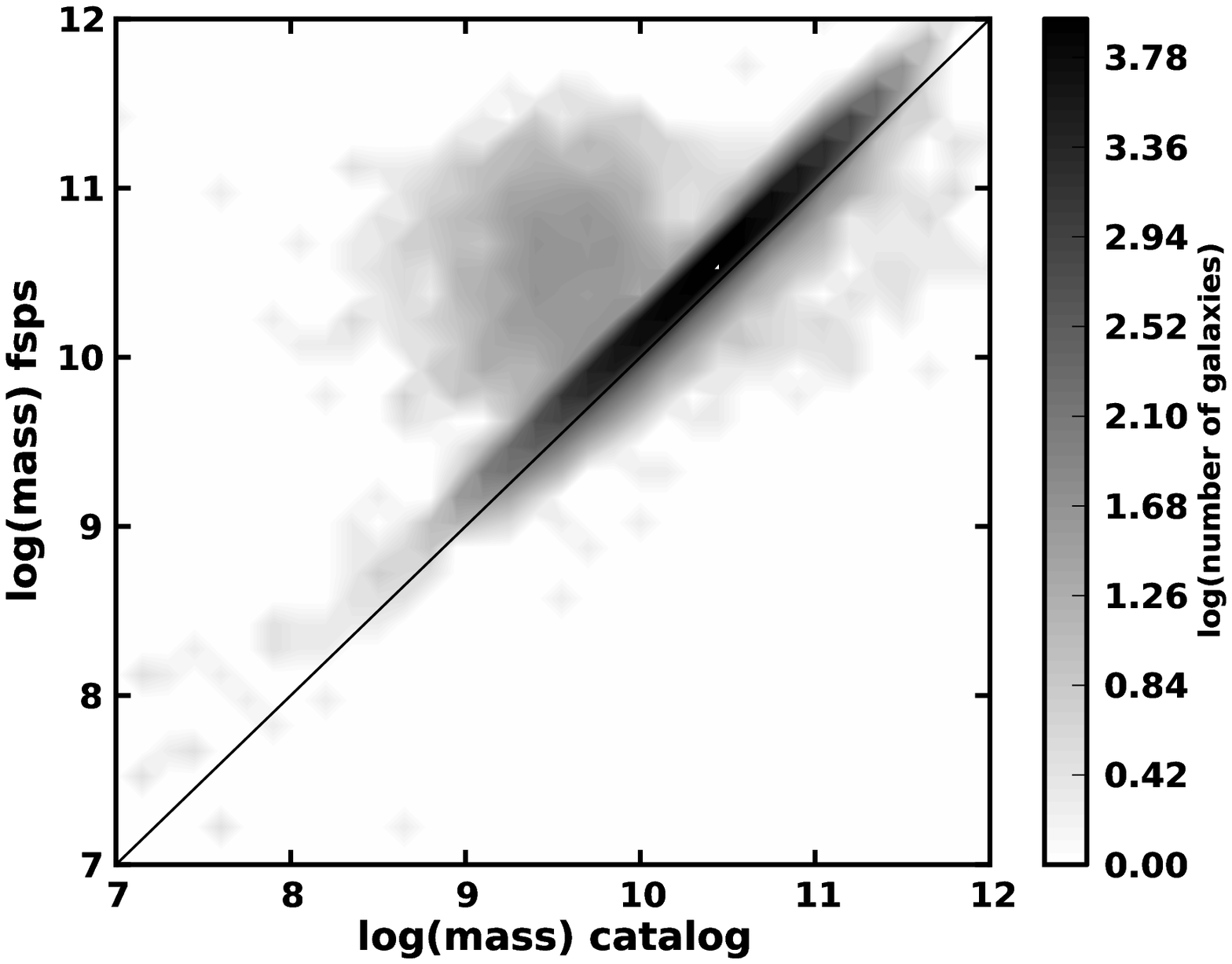} & \includegraphics*[scale=0.4]{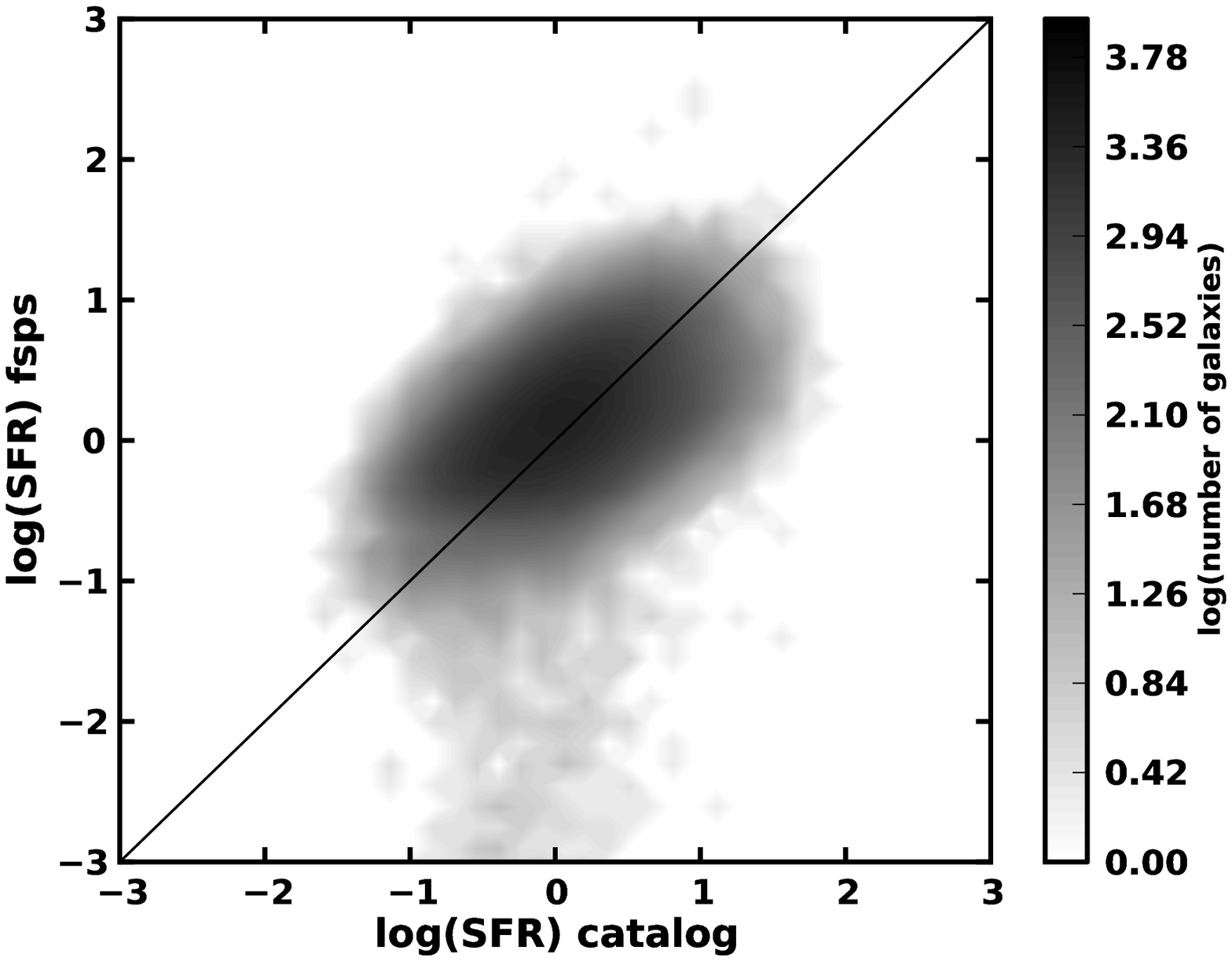}
\end{tabular}
\caption{Density plots (greyscale levels in log space) of the G11 fits compared to the catalog mass and the SFR measured from H$_\alpha$. \textit{Left:} the G11 stellar mass estimates against the SDSS DR-7 mass estimates from \citet{sal07}. There is also a slight offset in these values, possibly due to the use of a different IMF. \textit{Right:} the G11 SFR values against the values measured directly from H$_\alpha$ emission lines. \label{clouds_fsps}}
\end{figure*}

Figure \ref{clouds_fsps} shows the mass and SFR values from the G11 fitting method against those from the catalog. While the mass values show a small offset similar to the offset from Z-PEG, this offset is likely due to the use of a different IMF in the G11 fitting method. The SFR values show a tendency to have less spread, and with a less-steep slope. Slightly different from Z-PEG, G11 does not generate zero SFR objects. The objects that extend down to low SFR values are likely the same objects that result in zero SFR galaxies in the Z-PEG fits. There is also a curious group of galaxies in stellar mass space, with much larger values than the catalog. However, the number of galaxies in this region is relatively small compared to the full sample, so this does not systematically affect the FMR. The greyscale levels are drawn in log space, so the number of galaxies in this region is very small compared to the main sample.

\begin{figure*}
\centering
\begin{tabular}{cc}
\includegraphics*[scale=0.4]{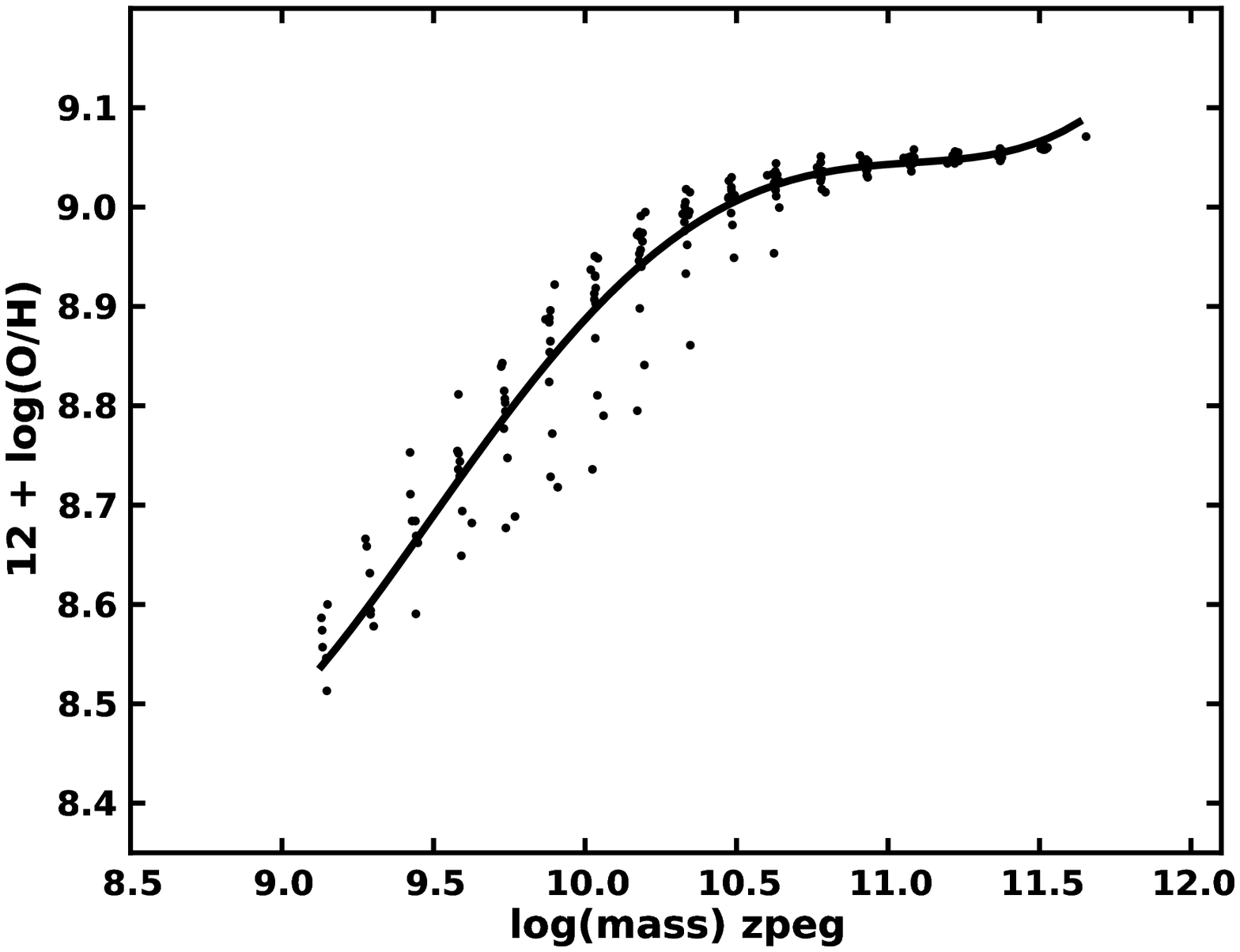} & \includegraphics*[scale=0.4]{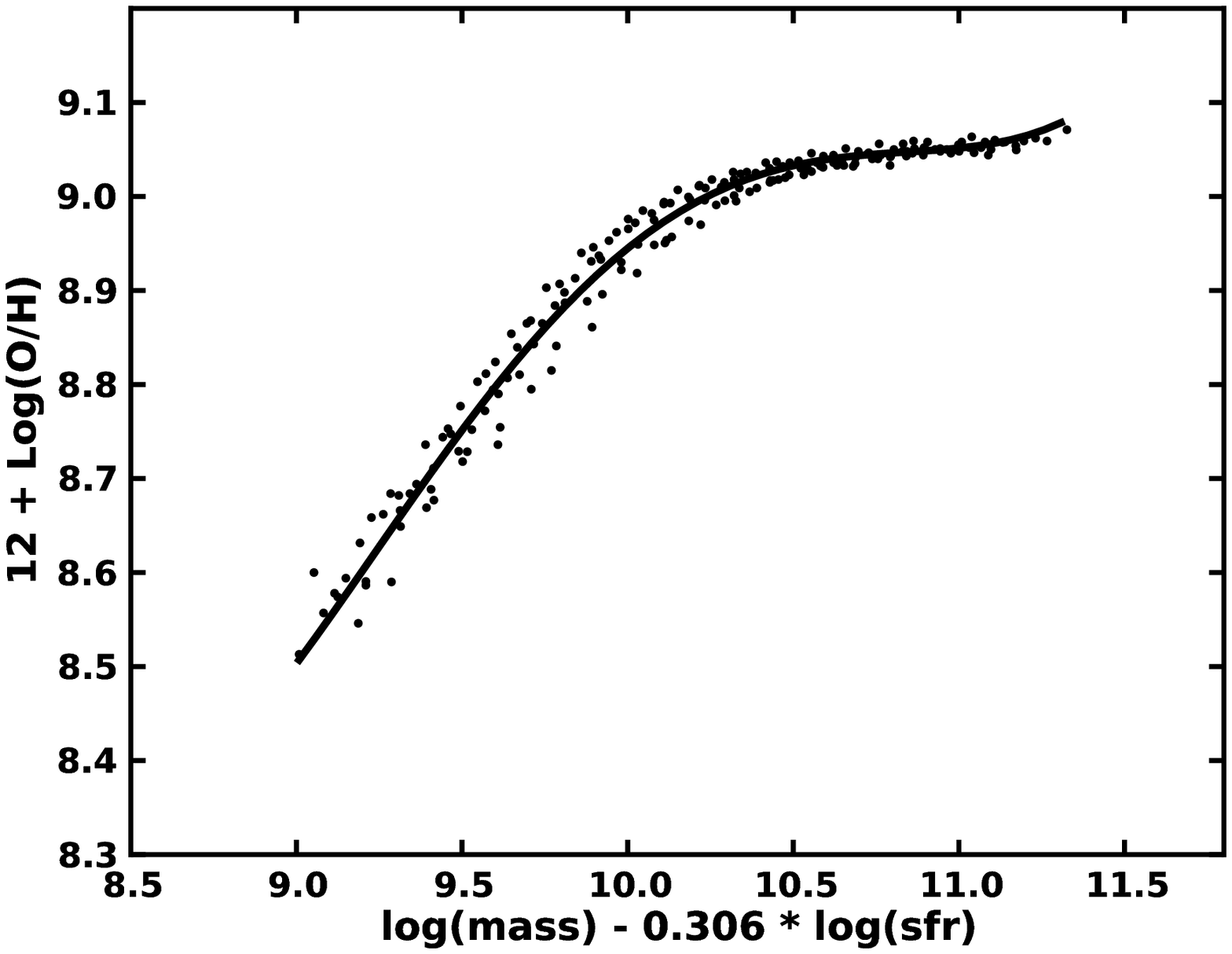}
\end{tabular}
\caption{The fundamental metallicity relation using Z-PEG applied to broadband photometry. Adding a portion of the SFR significantly reduces the scatter in the metallicity estimate. All mass and SFR values were estimated photometrically by Z-PEG, from ugriz photometry from SDSS-DR7. This demonstrates the power of photometric fitting of multi-band photometry to estimate the metallicity of a large sample of galaxies. \textit{Left:} the metallicity of the SDSS-DR7 galaxies as a function of the Z-PEG estimate of the galaxy stellar mass. The points represent the median of the metallicity and the mean of the mass, in bins 0.15 dex wide in the mass.  \textit{Right:} The same objects, also with 0.15 dex wide bins in the Z-PEG estimated SFR, showing the reduction in scatter that the SFR provides. \label{fmr}}
\end{figure*}

We use the same binning method as in section \ref{man_fmr} to calculate the median metallicity values as measured from the emission lines. Using Z-PEG to fit the ugriz multi-band photometry for the 140,987 SDSS galaxies passing our cuts, we reproduce the reduction in scatter as a result of combining mass and SFR to estimate metallicity. Figure \ref{fmr} shows the median metallicity as a function of mass (i.e. the Tremonti relation of \citealt{tre04}), and as a function of log(mass) $-$ 0.306 $\times$ log(SFR). The FMR reduces the dispersion from 0.05 in the mass-only case to 0.019. The equation of the best-fit line to the median metallicities is: 
\begin{equation}
\mathrm{12 + Log(O/H)} = 8.94 + 0.284\mathrm{x} - 0.226\mathrm{x}^2 - 0.010\mathrm{x}^3 + 0.058\mathrm{x}^4\label{eqn_ffmr}
\end{equation}
with x = $\mu_{\mathrm{FMR}}$ - 10, and $\mu_{\mathrm{FMR}}$ = log(mass) $-$ 0.306 $\times$ log(SFR).

\begin{figure*}
\centering
\begin{tabular}{cc}
\includegraphics*[scale=0.4]{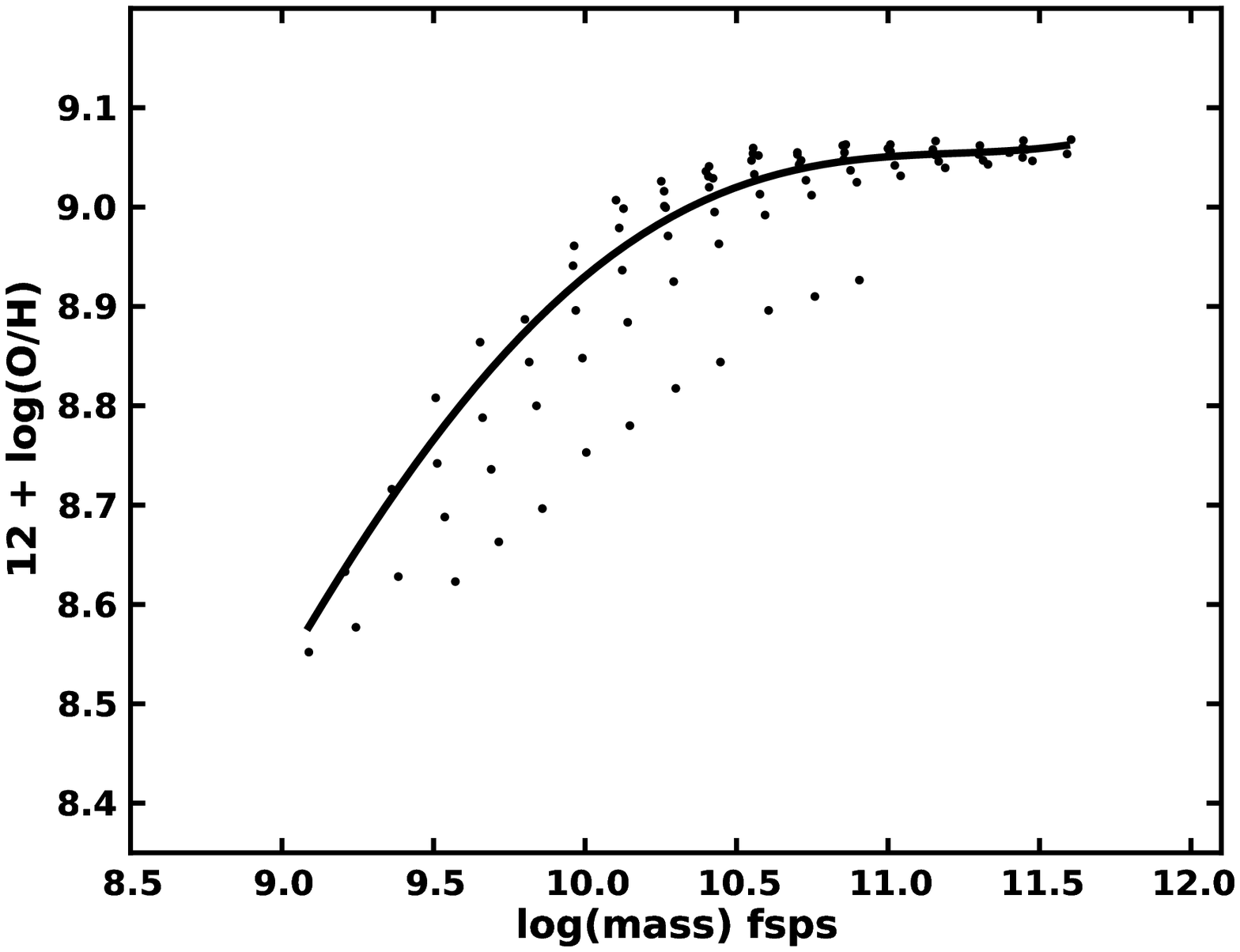} & \includegraphics*[scale=0.4]{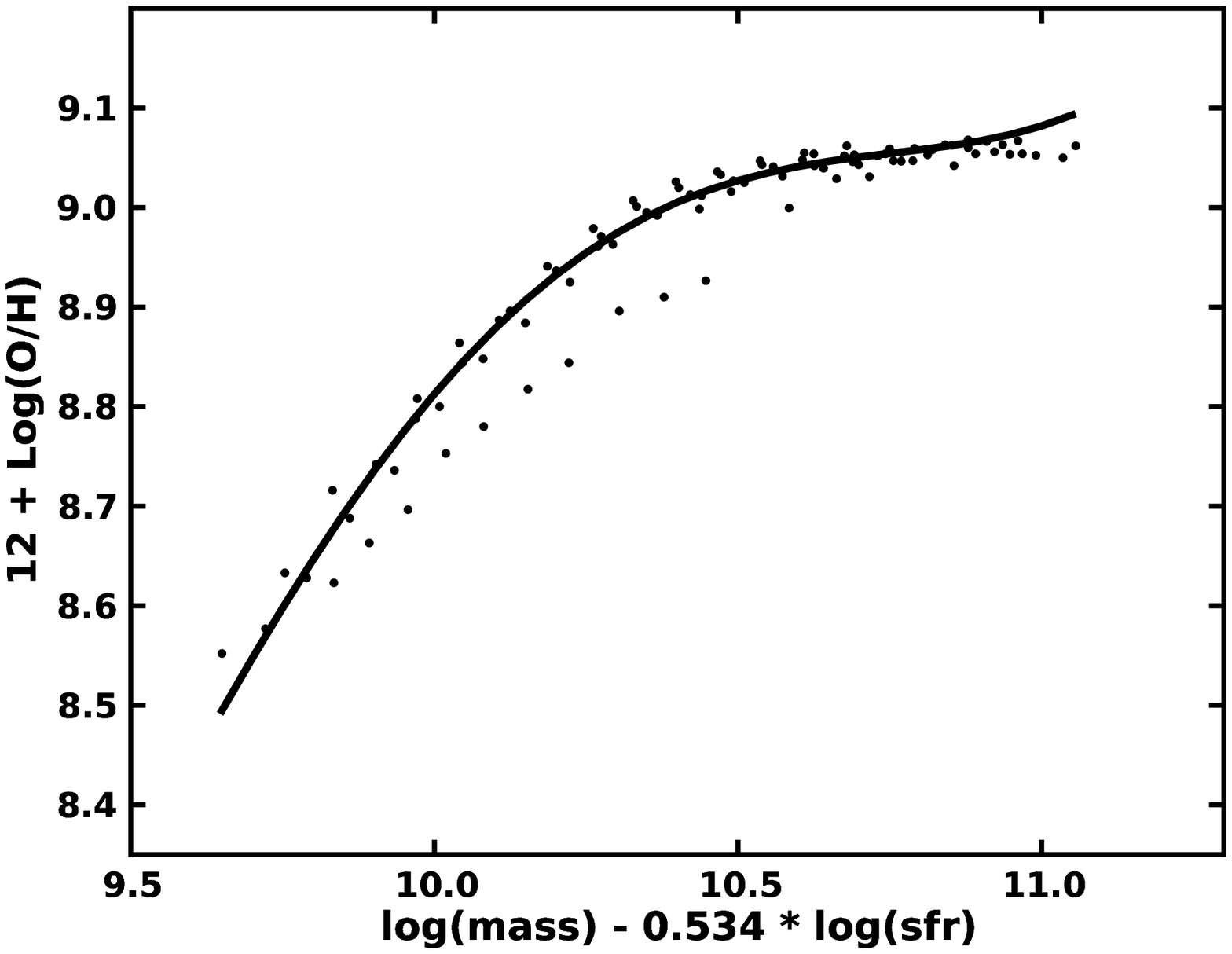}
\end{tabular}
\caption{The fundamental metallicity relation using the G11 fitting method. Adding a portion of the SFR significantly reduces the scatter in the metallicity estimate. All mass and SFR values were estimated photometrically using FSPS stellar models in the $\chi^2$ minimization routine of G11, from ugriz photometry from SDSS-DR7. This demonstrates the power of photometric fitting of multi-band photometry to estimate the metallicity of a large sample of galaxies, and corroborates our findings using Z-PEG. \textit{Left:} the metallicity of the SDSS-DR7 galaxies as a function of the G11 estimate of the galaxy mass. The points represent the median of the metallicity and the mean of the mass, in bins 0.15 dex wide in the mass.  \textit{Right:} The same objects, also with 0.15 dex wide bins in the G11 estimated SFR, showing the reduction in scatter that the SFR provides.   \label{fmr_fsps}}
\end{figure*}

Figure \ref{fmr_fsps} shows the correlation of Log(mass), and $\mu_{\mathrm{FMR}} =  \mathrm{Log(mass)} - 0.534 \times \mathrm{Log(SFR)}$ for G11 fits. Once again, the dispersion is significantly reduced, confirming the results of \citet{man10} and agreeing with the Z-PEG results in the sense that photometric fitting of mass and SFR values can reproduce the FMR. The stellar mass-only metallicity estimates have an RMS value of 0.065 dex, while the FMR estimated metallicity values have an RMS dispersion of 0.03 dex. The G11 FMR is given by: 
\begin{equation}
\mathrm{12 + Log(O/H)} = 8.81 + 0.718\mathrm{x} - 0.598\mathrm{x}^2 - 0.074\mathrm{x}^3 + 0.223\mathrm{x}^4
\end{equation}
with x = $\mu_{\mathrm{FMR}}$ - 10, and $\mu_{\mathrm{FMR}}$ = log(mass) - 0.534 $\times$ log(SFR). 

%Again, we caution that this is valid only for the metallicity region of our calibration sample, which for FSPS is 12 + Log(O/H) from 8.41 to 9.10.

\begin{figure*}
\centering
\begin{tabular}{cc}
\includegraphics*[scale=0.4]{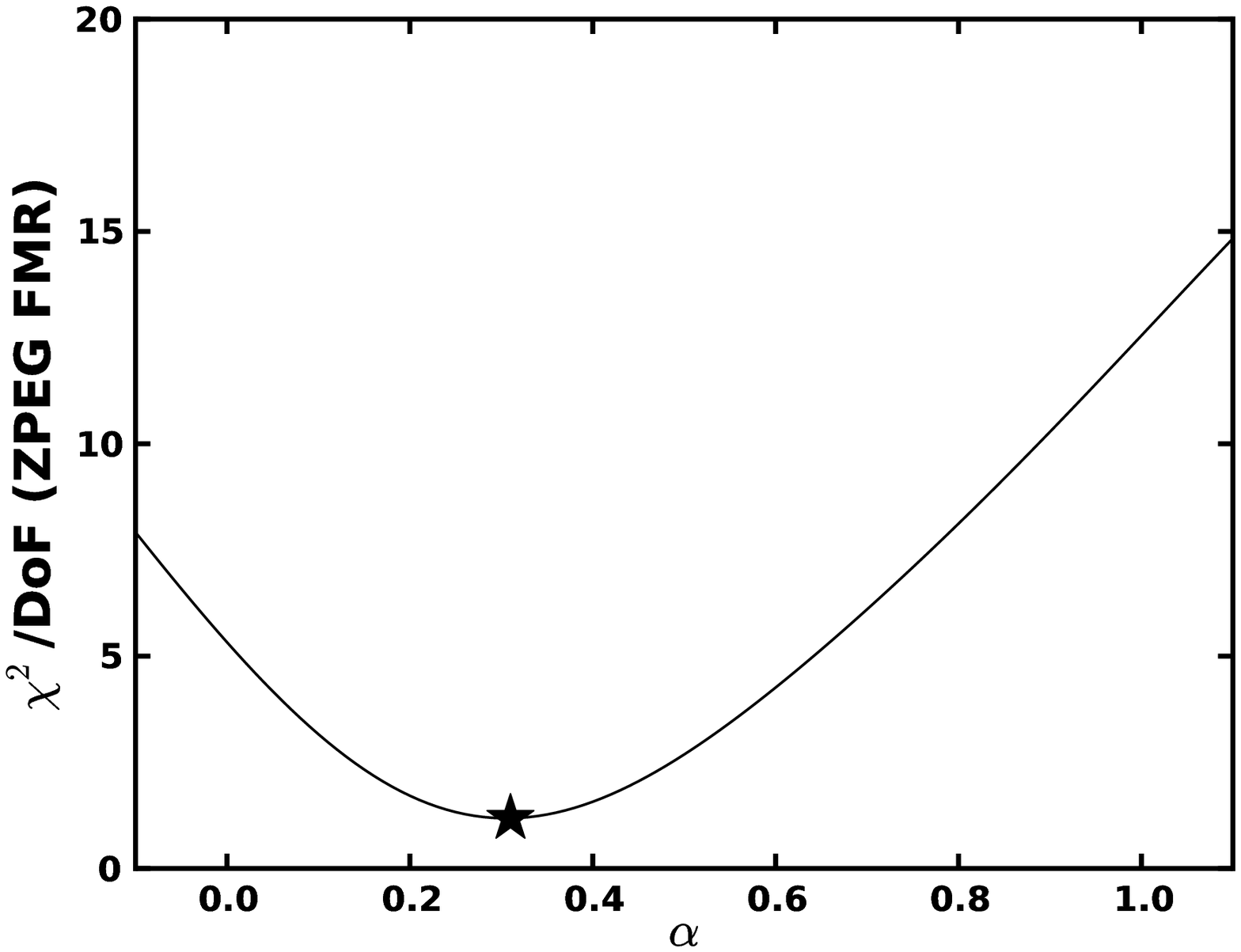} & \includegraphics*[scale=0.4]{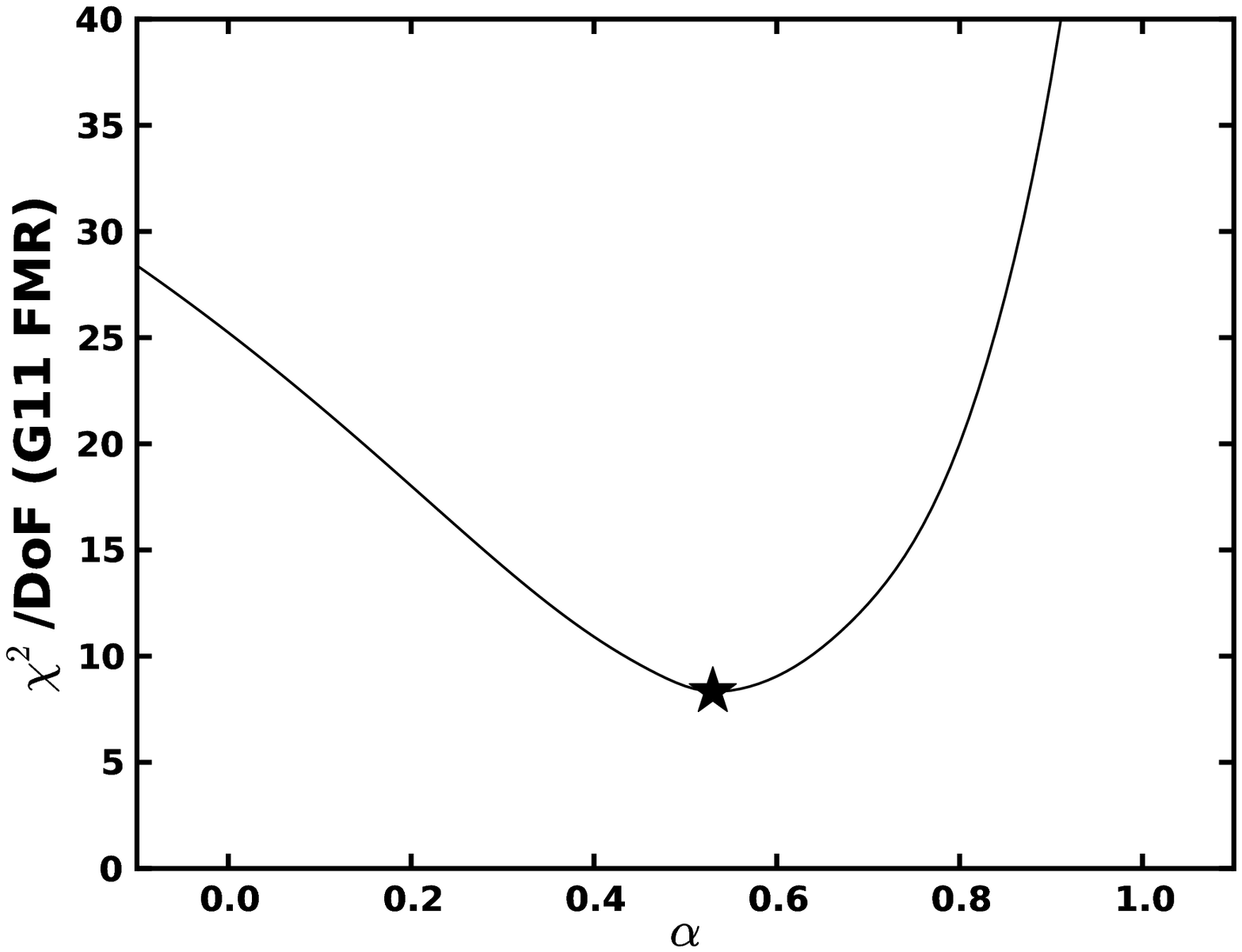}
\end{tabular}
\caption{The $\chi^2$ per degree of freedom value of the best-fit polynomial as a function of the parameter $\alpha$, from the Z-PEG FMR (\textit{left}), and the G11 FMR (\textit{right}) calibrations. The metallicity estimate is a function of log(mass) - $\alpha \times $log(SFR). For this plot we fix $\alpha$ at the indicated x-axis value and calculate the best fit fourth-order polynomial.  \label{chis}}
\end{figure*}

Figure \ref{chis} shows the dispersion in the metallicity as a function of the factor $\alpha$, where the fit parameter for the FMR is given by log(mass) $-$ $\alpha \times$log(SFR). The minimum $\chi^2$ value occurs where $\alpha$ = 0.306 for Z-PEG, and $\alpha$ = 0.534 for G11 fits. There is a difference in the FMR as calculated from G11 fits from \citet{man10} as well as our own analysis of the emission lines, which results in $\alpha$ around 0.32.  We discuss this issue in section \ref{sec:discussion}.

We reproduce the results of \citet{man10}, using photometric measurements alone, in that using both the stellar mass and a fraction of the SFR reduces the scatter in the metallicity estimates over the mass-metallicity relation of \citet{tre04}. The mass-only metallicity estimates for Z-PEG have an RMS dispersion of 0.05 dex, while the FMR estimates have a dispersion of 0.019 dex. For FSPS templates in the fitting method of G11,  the FMR reduces the dispersion from 0.06 to 0.03. This is compared to the RMS dispersion using the SDSS catalog of 0.02 dex, as in \citet{man10}. We have shown that photometrically estimated SFRs are sufficient to significantly reduce the scatter in the metallicity estimates using only multi-band photometry.  
%It is also important not to apply this relation to metallicities outside of the calibrated range, which in this case is 8.42 to 9.08. The fourth-order fit to the metallicities should not be used for metallicities outside this region, because it is unconstrained by the data. At the high metallicity end of the distribution, a linear extrapolation is likely safe. Other methods \citep{man11} have shown that linear extrapolations may be accurate towards lower metallicities as well.

\subsection{Applying the FMR to Hubble Residuals}
In this section we discuss the application of the FMR to the Hubble residuals of the SDSS-II SNe Ia in our sample. We present the analysis for our Z-PEG fits, and in section \ref{sec:discussion} we will present the same analysis for G11 fits. As a basic outline, we use two primary methods to compare the mass correlation with the FMR. First, we use both the mass-metallicity relation and the FMR to convert the SDSS-II host galaxy photometric measurements to metallicity. We then determine the correlation between these metallicities and the Hubble residuals, and compare them. Then, we compare the Hubble residual correlation of mass alone with the Hubble residual correlation of \mufmr, allowing $\alpha$ to vary. 

%In order to investigate the correlation between Hubble residuals and host galaxy mass, confirmed in \citet{kel10}, \citet{lam10}, \citet{sul10}, and \citet{gup11}, we test Hubble residual correlations with the metallicity parameter $\mu_{\mathrm{FMR}}$ = log(mass) $-$ $\alpha \times$ log(SFR), estimated using the Z-PEG and FSPS model photometric fits. Starting with the sample of 206 type Ia SN host galaxies from \citet{gup11}, we remove galaxies that have log(SFR) < -1.8 from the Z-PEG measurement. This leaves 193 host galaxies in the Z-PEG sample.  For \citet{gup11} fits, we remove galaxies with log(SFR) < -2, leaving 163 host galaxies. The primary difference between the Z-PEG fits and those using the \citet{gup11} fits is in the age constraint, and the difference in the number of galaxies passing these cuts shows the effect of this constraint on the SFR. This is discussed further in section \ref{sec:discussion}.
%and galaxies with metallicity values outside the Z-PEG calibrated FMR, which ranges from 12 + Log(O/H) of 8.434 to 9.086. This leaves 156 host galaxies to be tested in the Z-PEG sample. For our photometric fits using FSPS templates in the $\chi^2$ minimization process from \citet{gup11}, we remove host galaxies with Log(SFR) values less than -3, and also host galaxies with FMR metallicities not between 8.41 and 9.10. This leaves 150 galaxies in the FSPS sample.

To fit a line to the Hubble residuals as a function of any parameter, we use the Monte Carlo Markov Chain (MCMC) method LINMIX \citep{kel07}. This allows for a robust measurement of the linear fit parameters while taking into account the error in the independent variable. Similar to \citet{gup11}, we use 100,000 MCMC realizations, and LINMIX provides the posterior distribution of the slope and intercept values of the best-fit line. Using these distributions, which are highly gaussian, we fit a normal distribution to each of the resulting posterior distributions for the slope and intercept. The mean of the gaussian is considered the best-fit value (for either the slope or the intercept), and the $\sigma$ of the gaussian is considered the error. Dividing the mean by this $\sigma$ gives an indication of the statistical significance from zero slope, in terms of the number of standard deviations. 

While the mean and $\sigma$ values of our gaussian fits determine the statistical significance of any potential non-zero slope, this is not a quantitative measurement of the goodness-of-fit. To estimate how well the line fits the data we use the $\chi^2$ statistic. This allows us to directly compare multiple different independent variables based on the resulting $\chi^2$ value. In calculating the $\chi^2$ value of the best-fit line, we use both dependent and independent error sources; for the independent variable errors, we propagate the error by fixing the value of the FMR parameter $\alpha$ at the value of the best-fit FMR. As part of our analysis, we perform statistical testing as a function of $\alpha$ and fixing the value for error calculations prevents the error bars from varying with this parameter. 
%In calculating $\chi^2$, we do not include the independent variable errors. This is because for the FMR parameter $\mu_{\mathrm{FMR}}$, the errors are always larger than the errors on mass alone, and give an unfair advantage to $\mu_{\mathrm{FMR}}$ in any $\chi^2$ statistical testing. 

When comparing a Hubble residual fit of the mass alone to a fit using a version of $\mu_{\mathrm{FMR}}$, we use the likelihood ratio test (LRT). This statistic is used for nested models, in which the alternative model can be transformed into the base model by a simple restriction on the parameter space. The test statistic for the likelihood-ratio test, $\Lambda$, is given by:
\begin{equation}
\Lambda = -2\mathrm{ln}\left(\frac{P_0}{P_a}\right) \label{lrt_eqn}
\end{equation} 
where $P_0$ is the probability of the best-fit base model and $P_a$ is the probability of the best-fit alternative model. This test statistic obeys the $\chi^2$ distribution, with degrees of freedom $k = df_a - df_0$, where $df_a$ and $df_0$ are the degrees of freedom of the alternative and base models, respectively. The Neyman-Pearson lemma \citep{nplemma} indicates that this statistic has the highest power when distinguishing between two models. When we report a likelihood ratio test probability, it is the result of using this parameter $\Lambda$ as the test statistic in the $\chi^2$ distribution. Usage of this statistic in comparing the mass alone with the FMR is a test of the effect of the FMR parameter $\alpha$ on the Hubble residuals. In this scenario, a LRT probability less than 5\% indicates a statistical preference for the alternative model (the FMR). In all of our tests we use a version of the FMR as the alternative model, and the mass alone as the base model. The base model is a simple restriction of the alternative model in that mass alone is simply the FMR parameter $\mu_{\mathrm{FMR}}$ with $\alpha$ = 0. If $\alpha$ and therefore the FMR does not improve the goodness-of-fit of the best-fit line to the Hubble residuals, the LRT will return a value that is not statistically significant. 

Using our Z-PEG fits, we remove host galaxies with log(SFR) $<$ -1.8, leaving 193 galaxies from the original 206. Then we remove galaxies which have $\mu_{\mathrm{FMR}}$ values that fall outside the region of our photometrically calibrated FMR. This leaves 182 galaxies in the Z-PEG sample. For these galaxies left in our sample, we convert the photometrically measured stellar mass and SFR values using both the Z-PEG mass-metallicity relation and Z-PEG FMR. We then used LINMIX to fit the Hubble residuals as a function of both of these metallicities. In this analysis, we do not include the traditional 0.14 magnitude systematic error in the Hubble residuals. This error is intended to account for unknown sources of systematic error, and it is an unknown source of systematic error that we are testing. This reasoning is similar to that of \citet{gup11}.

Using the mass-metallicity relation values, we obtain a best-fit line with $\chi^2$ value of 481.08 with 180 degrees of freedom. This best-fit line has a non-zero slope with significance of 3.18$\sigma$.  Using the FMR metallicity values, we obtain a best-fit line with $\chi^2$ value of 468.5 with 179 degrees of freedom. This non-zero slope has a significance of 3.10$\sigma$. Using the likelihood ratio test to compare these fits, with the FMR parameter $\alpha$ as the lone degree of freedom, produces a LRT statistic $\lambda$ of 6.98 with one degree of freedom. In the $\chi^2$ distribution this results in a probability value of 0.008, indicating that the improvement in $\chi^2$ due to the addition of the FMR parameter $\alpha$ is significant at greater than 99\% confidence. 

\begin{figure*}
\centering
\begin{tabular}{cc}
\includegraphics*[scale=0.4]{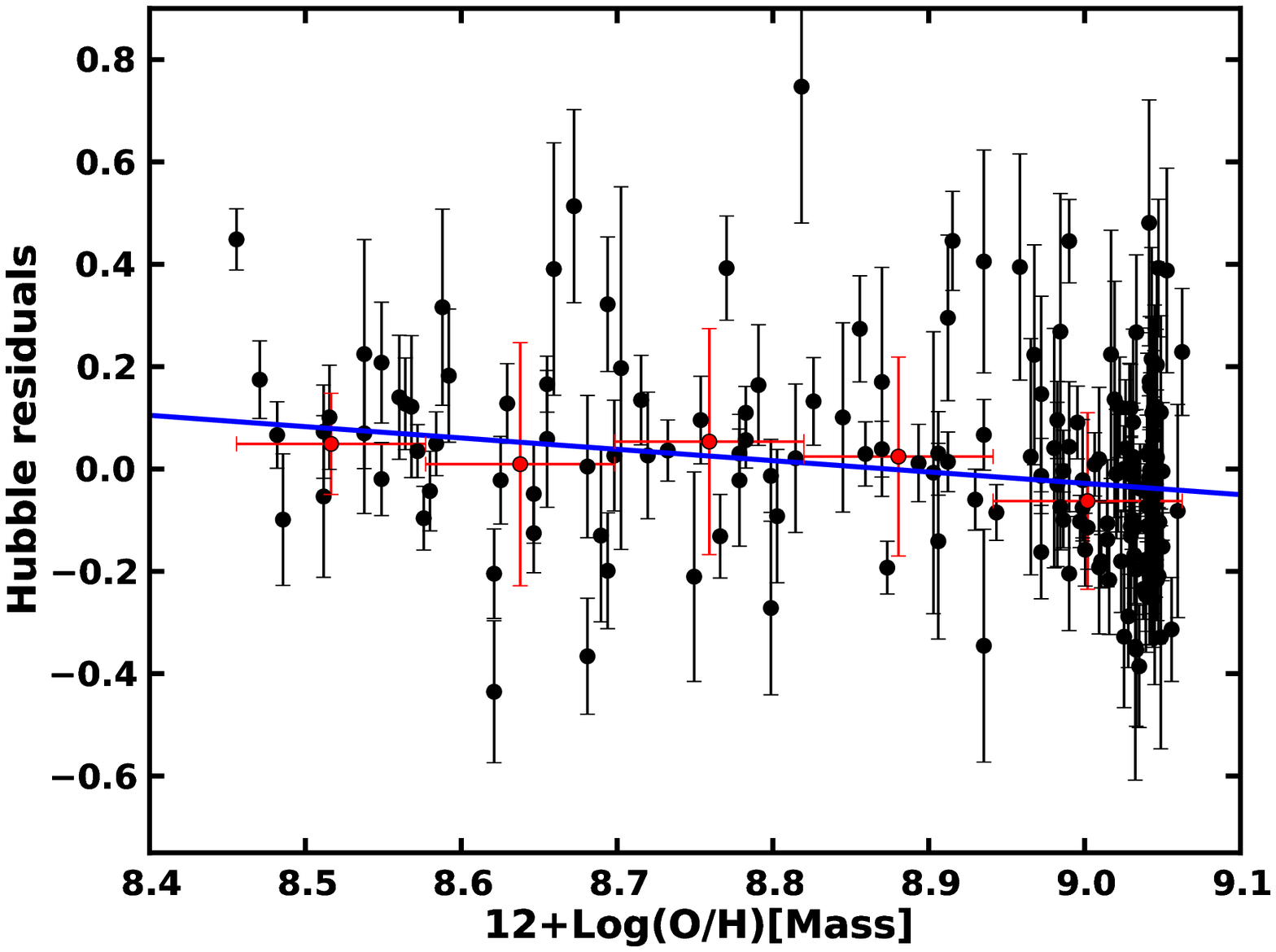} & \includegraphics*[scale=0.4]{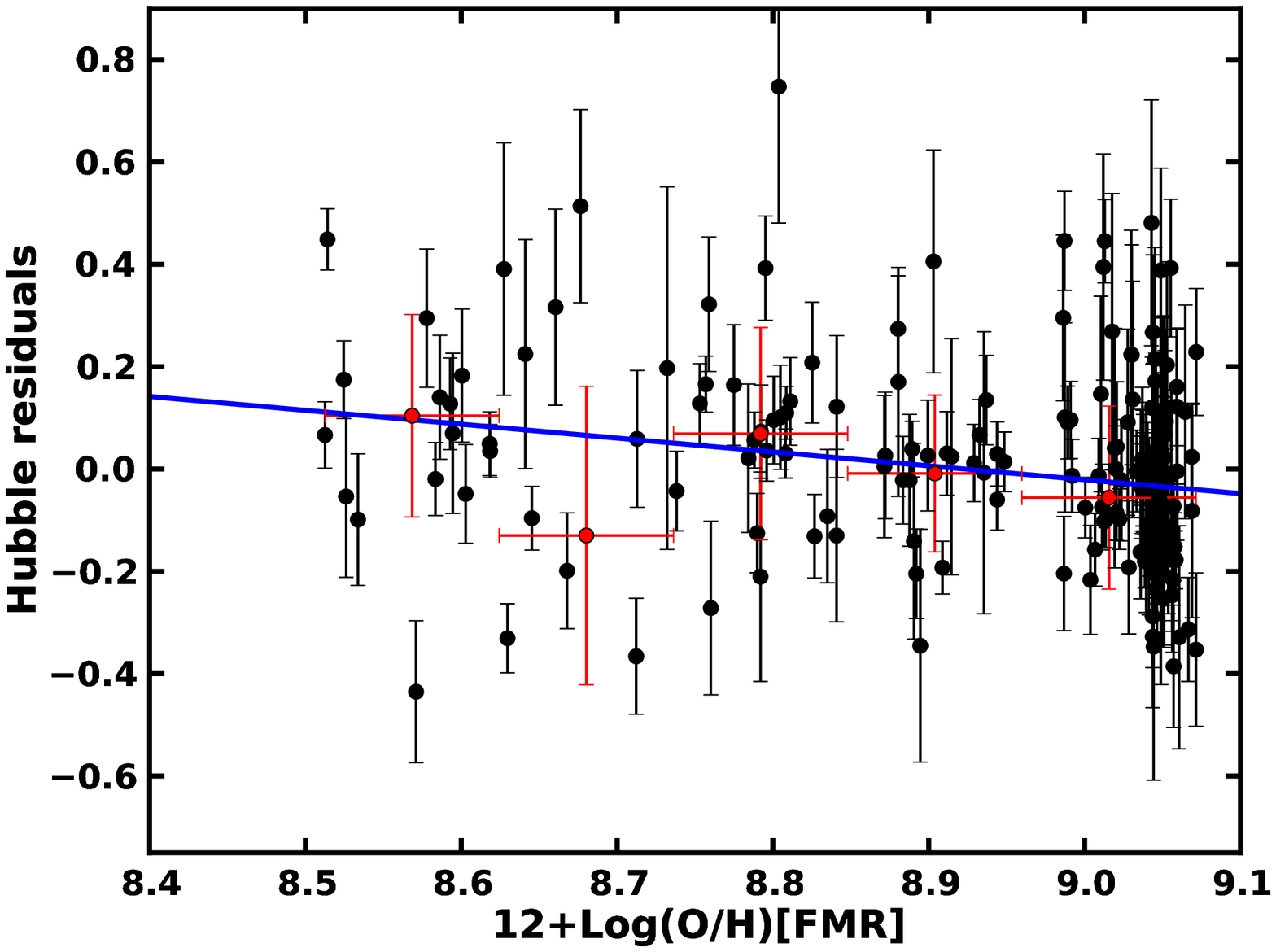}
\end{tabular}
\caption{The Hubble residuals as a function of metallicity converted from mass (left), and the fundamental metallicity relation (right) for Z-PEG.  The blue line is the linear fit to the Hubble residuals, and the red dots are the median values of the Hubble residuals. The likelihood-ratio test shows that the $\chi^2$ value of the best fit line with the FMR metallicities is significantly better than that of the mass-only values, at a significance of 99\%.  \label{hrmu}}
\end{figure*}

The fit of the Hubble residuals versus mass-metallicity relation has a slope of $-0.222$ (0.070) with an intercept of 1.966(0.620). The fit with respect to the FMR metallicity has a slope of $-0.271$ (0.087) with an intercept of 2.420 (0.782). Figure \ref{hrmu} shows the Hubble residuals versus the metallicity estimates for both scenarios. The metallicity reaches a maximum near 9.07, and many galaxies are grouped in this region. The inclusion of part of the SFR in the metallicity calculation tends to move points left to right in these plots, and the LRT indicates that the fit to the FMR values is statistically preferred. At first glance, it seems odd that two fits that both have a 3$\sigma$ significance of non-zero slope would produce a LRT probability that prefers one over the other. It is important to remember that the significance of the non-zero slope is in no way an estimate of the goodness-of-fit. It only contains information about whether or not there is a correlation between the dependent and independent variables being tested. It cannot necessarily provide information about which independent variable is better at reducing the scatter in the distribution.

The significance of our result is two-fold. First, we have shown that the Hubble residuals of 182 SDSS-II are strongly correlated with the metallicity estimates from both the mass and the FMR. We have also shown that the FMR provides a statistically better fit to the Hubble residuals, an indication that the stellar mass correlation with Hubble residuals is a tracer of a metallicity correlation. 

%RE-WRITE FROM HERE: Discuss direct comparison of metallicity measurements from mass and FMR. 

%For the 193 SDSS-II host galaxies in the Z-PEG sample, we performed MCMC linear fitting of the Hubble residuals as a function of mass, and the FMR parameter $\mu_{\mathrm{FMR}}$. For the Hubble residuals, we do not include the traditional 0.14 magnitude systematic error in quadrature, similar to \citet{gup11}. We are studying potential sources of systematic error, and adding 0.14 magnitudes in quadrature will reduce the impact of any sources that we study. The $\chi^2$ values for Hubble residuals versus mass and $\mu_{\mathrm{FMR}}$ are 409 and 399 with 154 degrees of freedom (NEW NUMBERS WHEN LRT RUN IS FINISHED). The fits can be seen in figure \ref{hrmu}. Before fitting any of these parameters the Hubble residuals have a $\chi^2$ value of 428 with 156 degrees of freedom (NEW NUMBER). The statistical significance of the correlation in terms of the number of standard deviations from zero slope are: Log(mass), 2.01$\sigma$, $\mu_{\mathrm{FMR}}$, 1.65$\sigma$. 
%The statistical significance of $\mu_{\mathrm{FMR}}$ is lower because in the MCMC fit, it has larger independent error values. 

\begin{figure*}
\centering
\includegraphics*[scale=0.7]{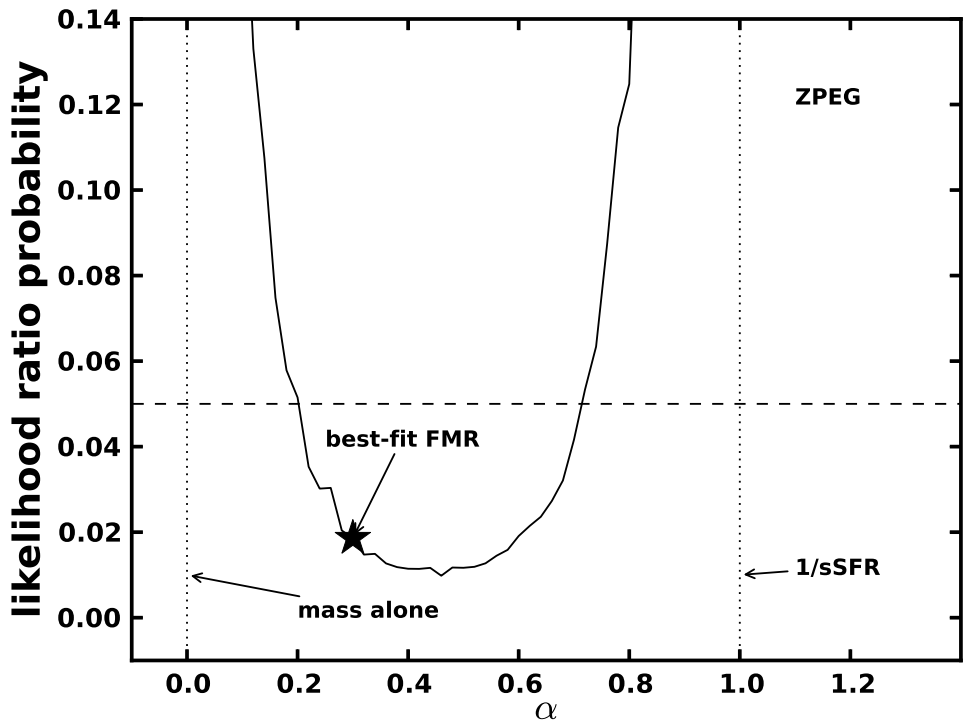}
\caption{The likelihood ratio test probability as a function of the FMR parameter $\alpha$, for Z-PEG. The probability value is a result of using the likelihood ratio test statistic $\Lambda$, from equation \ref{lrt_eqn}, as the test statistic in the $\chi^2$ probability distribution with 1 degree of freedom.ÊThis compares each value of $\alpha$ against the mass alone (shown by a vertical line at $\alpha$=0). This minimization was performed using the Hubble residuals, and is mathematically independent of the FMR. The region of parameter space where the probability is below 5\% agrees with the best-fit $\alpha$ value of the independently measured FMR (indicated by a star). This indicates that a linear combination of the stellar mass and SFR improves the Hubble residuals, and implies that the stellar mass correlation with SN Ia distances is tracing a correlation with the metallicity.    \label{lrt_zpeg}}
\end{figure*}

Because galaxies in the sample have a maximum metallicity value which causes galaxies to bunch up at high metallicity, we also test the Hubble residuals against the mass values, and against the \mufmr   values over a range of $\alpha$ values. Figure \ref{lrt_zpeg} illustrates the LRT probability value as a function of $\alpha$. To produce this plot, we fit the Hubble residuals versus $\mu_{\mathrm{FMR}}$ for $\alpha$ values ranging from $-0.1$ to 1.5. We compared each fit to the case where $\alpha$ = 0 using the $\chi^2$ value of the fit and the LRT. All errors were fixed to the $\alpha$ value for the best-fit FMR, simply to avoid the independent errors varying between fits. The fixed value of $\alpha$ in the error calculation does not affect these results. The vertical dotted line at $\alpha$ = 0 represents the mass alone, and as $\chi^2$ values are compared to this value, the LRT probability for $\alpha$ = 0 is 1. The vertical dotted line at $\alpha$ = 1 represents 1/sSFR, a parameter highly correlated with the age of the population. Probability values less than 5\% indicate a significant improvement in the goodness-of-fit for the FMR parameter, suggesting a rejection of the null model (mass alone) in favor of the alternative model (non-zero $\alpha$). The star indicates the $\alpha$ value of the best-fit FMR from the SDSS DR7 galaxies. The region of statistical preference in $\alpha$ agrees well with the independent formulation of the FMR, and does not favor the mass alone or the inverse of the specific star-formation rate. This is confirmation of the correlation of the Hubble residuals with metallicity, and further evidence that the Hubble residual correlation with stellar mass is tracing a correlation with the metallicity.

%For our LINMIX MCMC linear fits to the Hubble residuals in the FSPS sample, we obtain best-fit $\chi^2$ values of: Log(mass) = 456, $\mu_{\mathrm{FMR}} =$ 445 (NEW NUMBERS). These fits both have 148 (NEW NUMBER) degrees of freedom, and are shown in figure \ref{hrmu_fsps}. The rest of this section will discuss these numbers in greater depth.

We reiterate that since we are using the $\mu_{\mathrm{FMR}}$ value, log(mass) $-$ $\alpha$ $\times$ log(SFR), the minimization that is performed to produce figure \ref{lrt_zpeg} via the likelihood ratio test is independent from the minimization performed when calculating the FMR. The only links are the log(mass) and log(SFR) values. The results show a definite improvement in the Hubble residual correlation when using Log(mass) and a fraction of Log(SFR). While not definitive, this strongly suggests that the Hubble residual correlation with stellar mass is actually tracing a correlation with the metallicity, in the case of star-forming host galaxies.

\section{Discussion}
\label{sec:discussion}

\subsection{Comparing the Mannucci FMR to Z-PEG and \citet{gup11}}
\label{sec:fmrcompare}

We have demonstrated that the FMR can be formulated from broad band colors alone, but the parameters are dependent on the method used to estimate the stellar mass and SFR. In this section, we discuss some possible reasons for these differences in the functional form of the FMR. 

\begin{figure*}
\centering
\includegraphics*[scale=0.6]{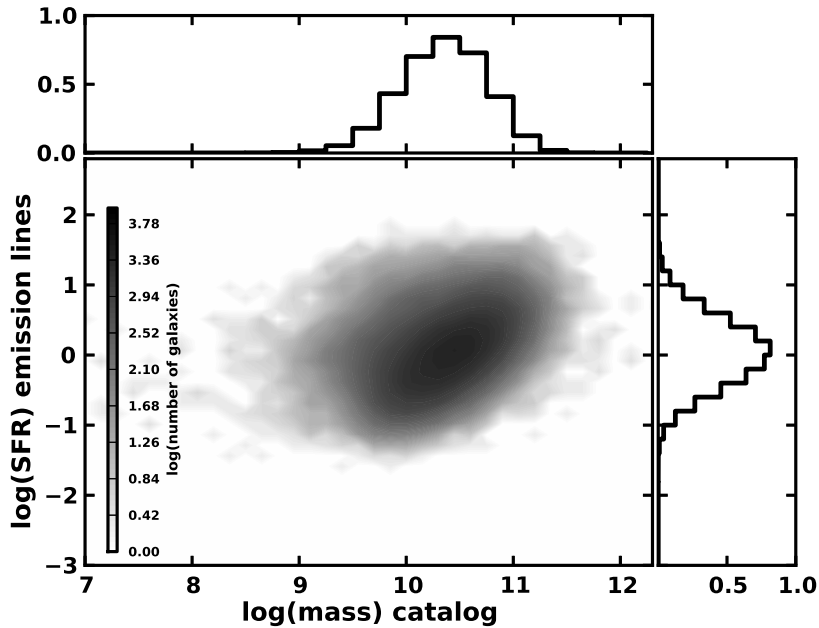} 
\caption{Greyscale density plot (contour levels in log space) of the sample of galaxies with mass values from the SDSS DR-7 catalogs, and SFR values measured from H$\alpha$.  \label{cloud_scat}}
\end{figure*}

\begin{figure*}
\centering
\includegraphics*[scale=0.6]{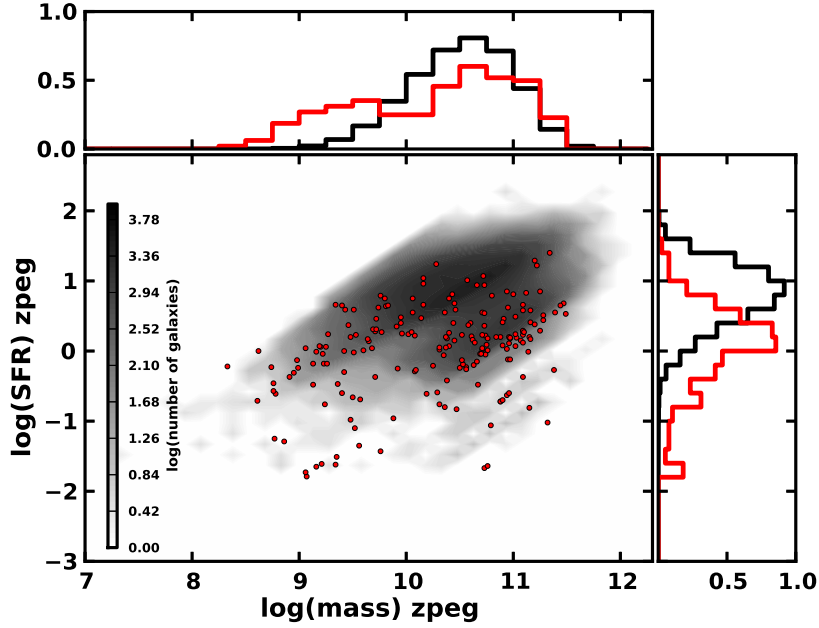} 
\caption{Greyscale density plot (contour levels in log space) of the sample of galaxies with mass and SFR values measured with Z-PEG. Shown in red are the SDSS-II hosts. The SFR distributions are different because of a bias towards high SFR galaxies due to our selection criteria in the SDSS DR7 catalog. \label{cloud_zpeg}}
\end{figure*}

\begin{figure*}
\centering
\includegraphics*[scale=0.6]{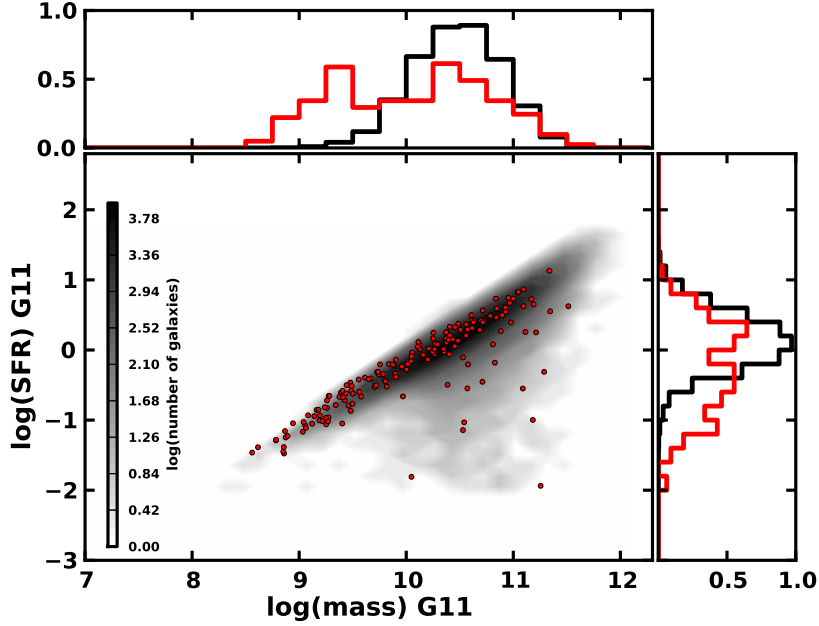} 
\caption{Greyscale density plot (contour levels in log space) of the sample of galaxies with mass and SFR values measured with FSPS. Shown in red are the SDSS-II hosts. \label{cloud_fsps}}
\end{figure*}

Figures \ref{cloud_scat}, \ref{cloud_zpeg}, and \ref{cloud_fsps}, contain some evidence as to the main difference between the estimated mass and SFR values that causes the fluctuation in $\alpha$. These figures show the log(SFR) vs log(mass) for the catalog, Z-PEG, and G11 model $\chi^2$ minimization techniques. The simplest difference to notice on first glance is the systematically lower SFR values from G11 as compared to Z-PEG. However, testing indicated that applying systematic offsets to the SFR values did not significantly change the best-fit $\alpha$ parameter in the FMR.

The next difference is the systematic offset in stellar mass values present in both Z-PEG and G11 as compared to the catalog. This effect can be tested by ``mixing and matching'' mass estimates in the FMR. In other words, using one measurement of the SFR with the three different measurements of the mass (catalog, Z-PEG, and G11 models). Doing so results in a spread of $\alpha$ values around 0.1 for each SFR method, not nearly enough to explain the difference in the FMR compared to the catalog stellar masses and SFR from H$\alpha$. We conclude that the mass offset is not the main source of $\alpha$ variation between FMR calibration sets. 

\begin{figure*}
\centering
\vspace{-88pt}
\includegraphics*[scale=0.8]{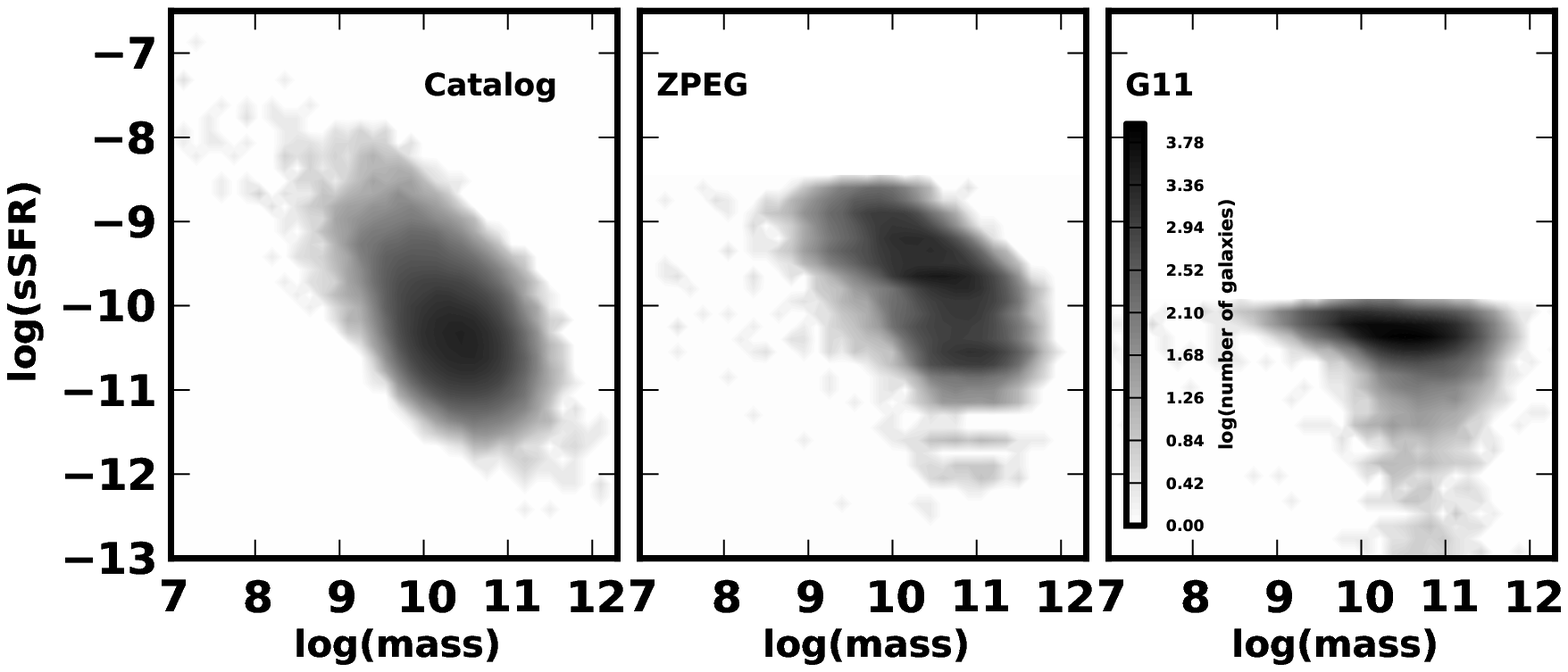}
\vspace{-60pt}
\caption{The specific star formation rate versus mass, for: \textit{left:} the DR7 catalogs, \textit{middle:} Z-PEG, and \textit{right:} the fits to FSPS templates. The cutoff and lack of spread in sSFR values in the photometric fitting techniques explain the difference in the FMR $\alpha$ value between the catalog and Z-PEG FMR formulations, which agree quite well, and the G11 formulation. \label{ssfr}}
\end{figure*}

The last major difference is in the specific star-formation rate. In figures  \ref{cloud_scat}, \ref{cloud_zpeg}, and \ref{cloud_fsps}, we show the SFR versus the mass. The distribution is much narrower for the G11 fits than for the catalog or Z-PEG values, indicating a lack of spread in the specific star formation rate. The sSFR against mass for all datasets is shown in figure \ref{ssfr}. The G11 fits show a marked difference in the sSFR distribution. G11 has no sSFR values above a certain level, which are clearly represented in the H$\alpha$ measured values. This is caused by an age constraint in the G11 fitting process combined with a declining SFR model. Indeed the spread in sSFR values measured from G11 is around half that of the H$\alpha$ measured values. This results in a significantly lower number of bins in the stellar mass / SFR plane, with the galaxies packed closer together in parameter space. This is the reason that the G11 FMR in figure \ref{fmr_fsps} has many fewer points than figure \ref{fmr_cat} and figure \ref{fmr}. In order to account for this mathematically, the minimization process compensates with a larger $\alpha$ value. Indeed ``mixing and matching'' the SFR with a fixed mass measurement method reproduces the large changes in $\alpha$ that we observe when using the different calibration methods to estimate the FMR. The small range in log(SFR) and log(sSFR) estimated from the photometric fits in G11 means that doubling $\alpha$ is required to get the same metallicity as measured directly from spectroscopy.

\subsection{\citet{gup11} Likelihood Ratio Test Results }
In this section we briefly reproduce the primary figures and results from the analysis section using G11 fits. Starting with the 206 host galaxies, removing galaxies with log(SFR) $<$ -2 leaves 163 galaxies. Removing galaxies that do not fall in the calibrated FMR region further reduces the sample to 135 galaxies. The primary reason for the difference in the number of galaxies is the specific star formation rate distribution of the fits; this was discussed in section \ref{sec:fmrcompare}. 

\begin{figure*}
\centering
\begin{tabular}{cc}
\includegraphics*[scale=0.4]{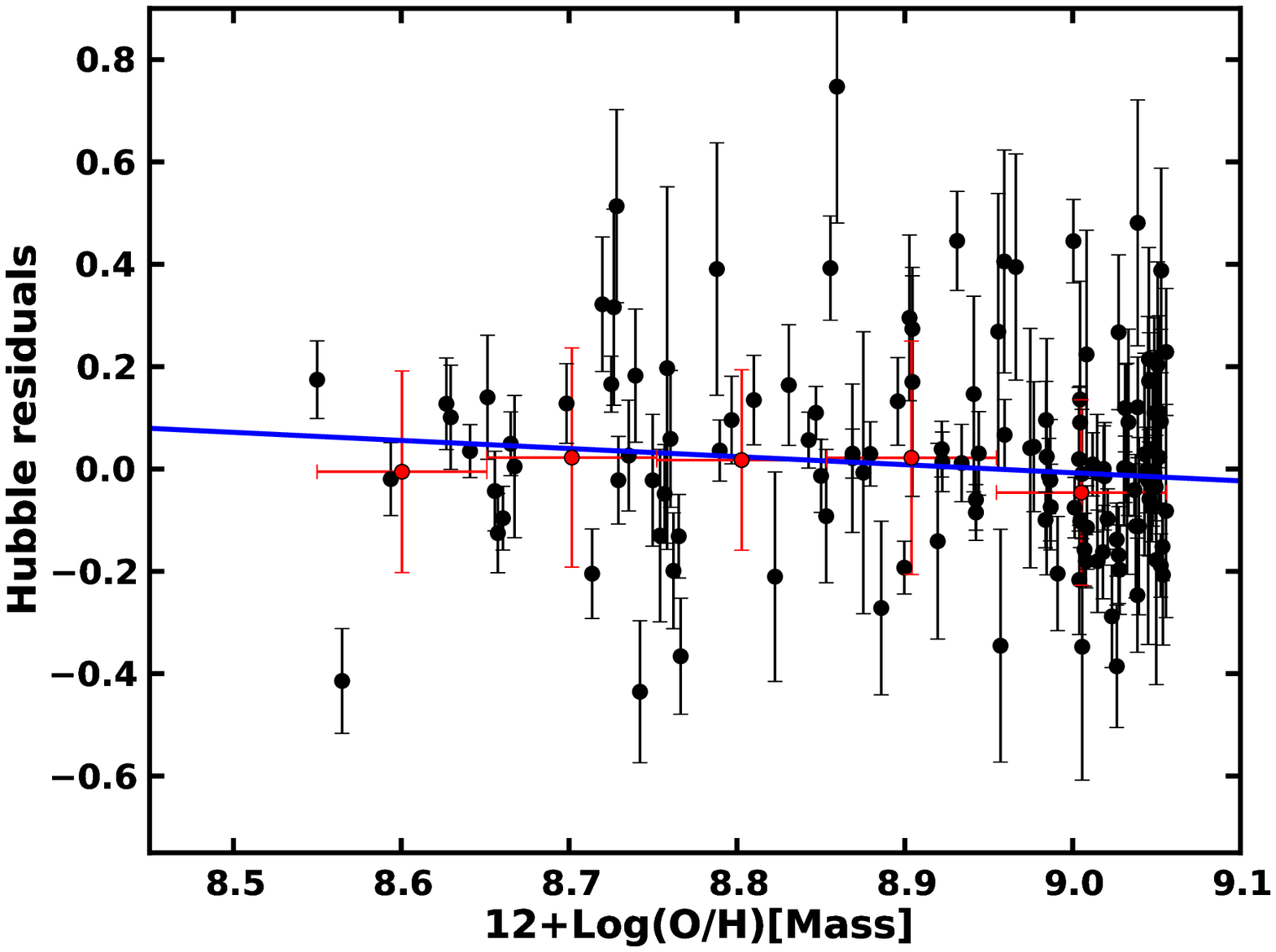} & \includegraphics*[scale=0.4]{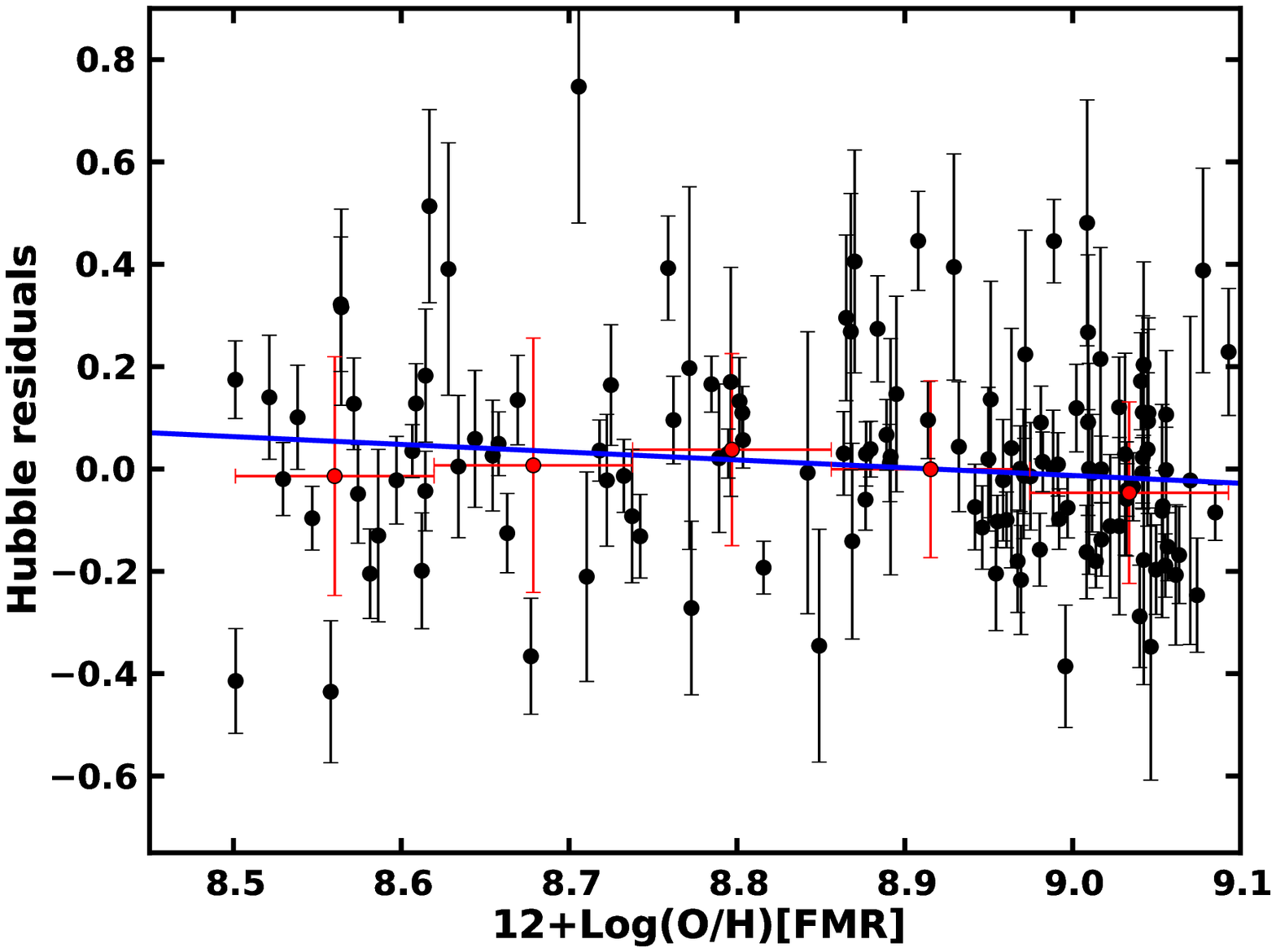}
\end{tabular}
\caption{The Hubble residuals as a function of metallicity converted from mass (left), and the fundamental metallicity relation (right) for G11.  The blue line is the linear fit to the Hubble residuals, and the red dots are the median values of the Hubble residuals. The likelihood-ratio test shows that the $\chi^2$ value of the best fit line with the FMR metallicities is not statistically preferred. Although the $\chi^2$ is lower, the significance of the improvement is less than 95\% confident. \label{hrmu_fsps}}
\end{figure*}

We first compare the best-fit $\chi^2$ value of the Hubble residuals versus mass-metallicity relation and the Hubble residuals versus FMR metallicity. The distributions can be seen in figure \ref{hrmu_fsps}. The mass-metallicity best-fit line has a $\chi^2$ value of 356.84 with 133 degrees of freedom. The FMR metallicity best-fit line has a $\chi^2$ value of 353.08 with 132 degrees of freedom. This results in a LRT statistic $\lambda$ = 1.39, with 1 degree of freedom. The probability of this value in the $\chi^2$ distribution is 0.239, a result that indicates no statistical preference for the FMR metallicity. 

\begin{figure*}
\centering
\includegraphics*[scale=0.7]{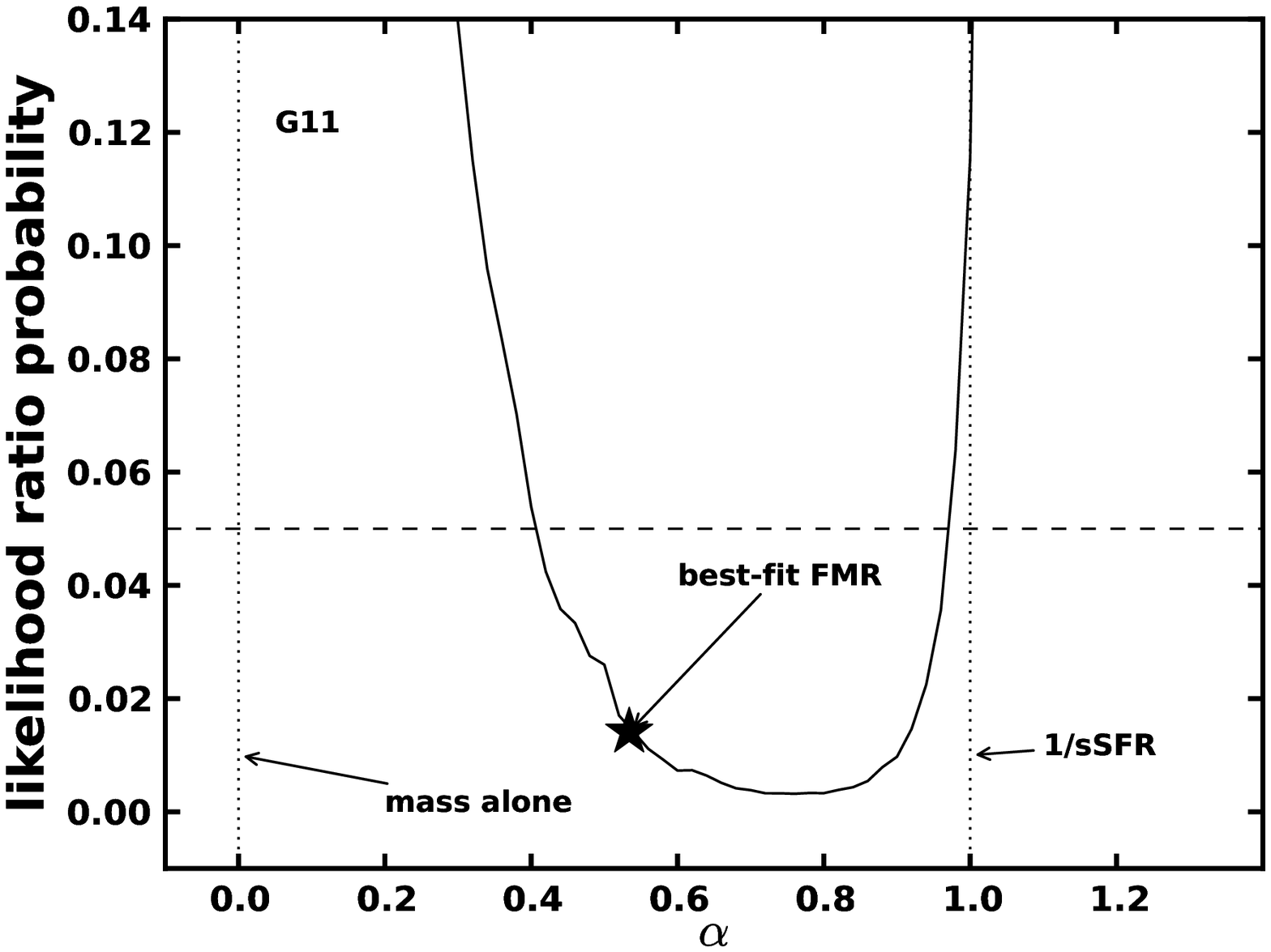}
\caption{The likelihood ratio test probability as a function of the FMR parameter $\alpha$, for G11. The probability value is a result of using the likelihood ratio test statistic $\Lambda$, from equation \ref{lrt_eqn}, as the test statistic in the $\chi^2$ probability distribution with 1 degree of freedom.ÊThis compares each value of $\alpha$ against the mass alone (shown by a vertical line at $\alpha$=0). This minimization was performed using the Hubble residuals, and is mathematically independent of the FMR. The region of parameter space where the probability is below 5\% agrees with the best-fit $\alpha$ value of the independently measured FMR (indicated by a star). This indicates the improvement of the FMR in reducing the scatter of the Hubble residuals and is evidence that the mass correlation is tracing the metallicity.    \label{lrt_fsps}}
\end{figure*}

The LRT as a function of the FMR parameter $\alpha$ using the mass and $\mu_{\mathrm{FMR}}$ values once again agrees with the FMR, in the sense that including a portion of the SFR reduces the Hubble residual scatter at a statistically significant level. Figure \ref{lrt_fsps} indicates that adding an element of the SFR to the mass reduces the Hubble residual scatter about the best-fit line, for $\alpha$ values that agree well with the G11 formulation of the FMR. This again confirms the metallicity correlation with Hubble residuals, and suggests that the Hubble residual correlation with mass is tracing this correlation with metallicity. 

As discussed in the previous section, the G11 fits remove many more galaxies from the sample due to the lack of spread in sSFR values. This lack of spread in sSFR is what causes the LRT test for direct comparison between metallicity estimates to be statistically insignificant. The loss of around 50 galaxies from the sample simply due to their mass and SFR values falling outside of the FMR calibration range indicates the importance of improving the way SED fitters estimate the SFR and therefore the recent star formation history. Similar to the default Z-PEG setup, G11 uses an age constraint which forces each template to contain old stars. This forces an unrealistic correlation between the mass and SFR values. While this constraint reduces the number of outliers in stellar mass, but greatly restricts parameter space with respect to the recent SFH. 

For our Z-PEG fits, relaxing the age constraint was a simple process. G11, however, uses FSPS templates that do not have a published set of parameters that mimic real galaxies (similar to \citealt{zpeg} for Z-PEG), so we were not able to reproduce the sSFR distribution of the catalog and Z-PEG values. Looking ahead for SED fitters and SN Ia studies alike, especially in light of the potential power of photometric fitting and the FMR, an improvement in the estimation of the SFR from photometric SED fitters is of paramount importance. The SFR values from SED fitting are highly dependent on the star-formation history used in creating the templates, and the constraints on the age of the template that are appropriate for stellar mass estimation are not generally appropriate for SFR estimation. Further study of photometric formulation of the FMR (for any purpose) will be highly dependent on improvements in the estimation of the SFR.

\subsection{Extending the FMR to Lower Metallicity}
While we have demonstrated the power of photometric fitters in improving metallicity estimates, and even Hubble residuals, our calibration sample was relatively low redshift. \citet{man10} claims no evolution of the FMR out to redshift 2.5, however, the number of galaxies beyond the SDSS sample, which only goes out to redshift 0.3, was small. As redshift increases, the mass decreases and the SFR increases, on average \citep{aasposter}, and so the value of the parameter $\mu_{\mathrm{FMR}}$ decreases as does the metallicity. It would be unwise to apply the FMR to galaxies that do not fall within the region of $\mu_{\mathrm{FMR}}$ and 12+Log(O/H) values that were used in the calibration sample. \citet{man11} was able to show that the FMR extends smoothly down to 12+Log(O/H) of around 8, using the lower-mass lower-metallicity SDSS galaxies that fell into bins that contained less than 50 galaxies. \citet{cre12} extends the FMR to about a redshift of 0.8, finding no evidence for evolution. This confirms the results in \citet{man10} and \citet{man11} with a higher level of significance. 

If the FMR is reliable out to redshift 2.5, as implied by \citet{man10} and \citet{cre12}, this would become a powerful tool for reliably estimating metallicities of a significant fraction of the universe. More work must be done to both extend the FMR to larger samples of low metallicity galaxies, and probe potential evolution of the FMR extending back in cosmic history. 

\section{Conclusions}
\label{sec:conclusions}

The fundamental metallicity relation shows that a combination of stellar mass and SFR improves metallicity estimates beyond those derived from stellar masses alone, i.e. the Tremonti relation \citep{tre04}. We have demonstrated the capability of two photometric methods, Z-PEG and $\chi^2$ minimization of FSPS models from \citet{gup11}, to reproduce the fundamental metallicity relation of \citet{man10}, and significantly reduce the scatter in metallicity estimates. This allows for accurate metallicity estimates based on multi-band photometry alone.

We have demonstrated the necessity of calibrating the FMR for the photometric fitter that will be used to estimate the stellar mass and SFR. In the case of \citet{gup11} fits, the relatively small variation in available specific SFR values greatly changes the calibration of the FMR. Using the Z-PEG or \citet{gup11} galaxy parameters with the specific formulation of the FMR from \citet{man10} would give results that were at best systematically biased in some way, and at worst nonsensical. 

We have calibrated the FMR for both the Z-PEG and G11 photometric fits, and fit a star-forming subsample of the 206 SDSS-II SN Ia host galaxies with these fitters in the SDSS ugriz bands. We have demonstrated that using the FMR parameter $\mu_{\mathrm{FMR}} = \mathrm{log(mass)} - \alpha \times \mathrm{log(SFR)}$  improves Hubble residual correlation beyond the stellar mass alone at a statistical significance of at least 99\%, using the likelihood ratio test. We note again that this sample contains only relatively high SFR galaxies. Still, the improvement in Hubble residual correlation suggests that the mass correlation confirmed by many studies is very likely tracing an underlying metallicity correlation. Generally, metallicity effects on SN Ia peak brightnesses should affect the amount of radioactive Nickel-56 produced, and therefore be corrected for by light-curve width corrections. If indeed the mass correlation traces a correlation with the metallicity, then this correlation is separate from the light-curve width correction and indicates other physical processes, possibly indicating that the Phillips relation itself is metallicity dependent \citep{kas09}. The calibration of the FMR for photometric fitters will allow ever greater precision in the measurement of cosmological distances using SNe Ia, as well as improving the quality of metallicity studies for galaxies where measuring spectra is difficult or nearly impossible. 

Future SN Ia searches will need to account for the correlation of SN Ia Hubble residuals with the FMR and metallicity illustrated in this work. Where possible, large samples of spectroscopic metallicity values for SN Ia host galaxies should be obtained. Multi-wavelength photometry that allows accurate estimation of the stellar mass and SFR of the host galaxies will be instrumental in using the FMR to correct SN Ia distances, where spectroscopic study is prohibitively expensive, as well as furthering our understanding of the SFR estimates from SED fitting. As SN Ia studies extend to high-redshift, the effect of host galaxy correlations on SN Ia distances can systematically change, and building on the work presented here will be a significant effort for the next generation of SN Ia surveys.


\begin{thebibliography}{}
\bibitem[Astier et al.(2006)]{ast06} Astier, P., Guy, J., Regnault, N., et al.\ 2006, \aap, 447, 31
\bibitem[Barris et al.(2004)]{bar04} Barris, B.~J., Tonry, J.~L., Blondin, S., et al.\ 2004, \apj, 602, 571
\bibitem[Brinchmann et al.(2004)]{brinch04} Brinchmann, J., Charlot, S., White, S.~D.~M., et al.\ 2004, \mnras, 351, 1151
\bibitem[Chabrier(2003)]{chab03} Chabrier, G.\ 2003, \pasp, 115, 763
\bibitem[Conley et al.(2011)]{con11} Conley, A., Guy, J., Sullivan, M., et al.\ 2011, \apjs, 192, 1
\bibitem[Conroy et al.(2009)]{fsps} Conroy, C., Gunn, J.~E., \& White, M.\ 2009, \apj, 699, 486
\bibitem[Conroy \& Gunn(2010)]{fsps2} Conroy, C., \& Gunn, J.~E.\ 2010, Astrophysics Source Code Library, 10043
\bibitem[Cresci et al.(2012)]{cre12} Cresci, G., Mannucci, F., Sommariva, V., et al.\ 2012, \mnras, 421, 262
\bibitem[D'Andrea et al.(2011)]{dandrea11} D'Andrea, C.~B., Gupta, R.~R., Sako, M., et al.\ 2011, \apj, 743, 172
\bibitem[Eisenstein et al.(2007)]{eisen07} Eisenstein, D.~J., Seo, H.-J., Sirko, E., \& Spergel, D.~N.\ 2007, \apj, 664, 675
\bibitem[Frieman et al.(2008)]{frie08} Frieman, J.~A., Bassett, B., Becker, A., et al.\ 2008, \aj, 135, 338
\bibitem[Gallagher et al.(2008)]{gal08} Gallagher, J.~S., Garnavich, P.~M., Caldwell, N., et al.\ 2008, \apj, 685, 752
\bibitem[Garnavich et al.(1998a)]{garn98a} Garnavich, P.~M., Kirshner, R.~P., Challis, P., et al.\ 1998, \apjl, 493, L53
\bibitem[Garnavich et al.(1998b)]{garn98b} Garnavich, P.~M., Jha, S., Challis, P., et al.\ 1998, \apj, 509, 74
\bibitem[Gunn et al.(1998)]{gunn98} Gunn, J.~E., Carr, M., Rockosi, C., et al.\ 1998, \aj, 116, 3040
\bibitem[Gunn et al.(2006)]{gunn06} Gunn, J.~E., Siegmund, W.~A., Mannery, E.~J., et al.\ 2006, \aj, 131, 2332
\bibitem[Gupta et al.(2011)]{gup11} Gupta, R.~R., D'Andrea, C.~B., Sako, M., et al.\ 2011, \apj, 740, 92
\bibitem[Guy et al.(2007)]{guy07} Guy, J., Astier, P., Baumont, S., et al.\ 2007, \aap, 466, 11
\bibitem[Guy et al.(2010)]{guy10} Guy, J., Sullivan, M., Conley, A., et al.\ 2010, \aap, 523, A7 
\bibitem[Hayden et al.(2012)]{aasposter} Hayden, B., Garnavich, P., \& Survey, C.~S.\ 2012, American Astronomical Society Meeting Abstracts \#219, 219, \#429.01 
\bibitem[Holtzman et al.(2008)]{hol08} Holtzman, J.~A., Marriner, J., Kessler, R., et al.\ 2008, \aj, 136, 2306
\bibitem[Johansson et al.(2012)]{joh12} Johansson, J., Thomas, D., Pforr, J., et al.\ 2012, arXiv:1211.1386
\bibitem[Kasen et al.(2009)]{kas09} Kasen, D., R{\"o}pke, F.~K., \& Woosley, S.~E.\ 2009, \nat, 460, 869
\bibitem[Kauffmann et al.(2003a)]{kauf03a} Kauffmann, G., Heckman, T.~M., Tremonti, C., et al.\ 2003, \mnras, 346, 1055 
\bibitem[Kauffmann et al.(2003b)]{kauf03b} Kauffmann, G., Heckman, T.~M., White, S.~D.~M., et al.\ 2003, \mnras, 341, 33
\bibitem[Kelly(2007)]{kel07} Kelly, B.~C.\ 2007, \apj, 665, 1489
\bibitem[Kelly et al.(2010)]{kel10} Kelly, P.~L., Hicken, M., Burke, D.~L., Mandel, K.~S., \& Kirshner, R.~P.\ 2010, \apj, 715, 743
\bibitem[Kennicutt(1998)]{kenn98} Kennicutt, R.~C., Jr.\ 1998, \araa, 36, 189 
\bibitem[Kessler et al.(2009a)]{kess09} Kessler, R., Becker, A.~C., Cinabro, D., et al.\ 2009, \apjs, 185, 32
\bibitem[Kessler et al.(2009b)]{kess09b} Kessler, R., Bernstein, J.~P., Cinabro, D., et al.\ 2009, \pasp, 121, 1028
\bibitem[Knop et al.(2003)]{knop03} Knop, R.~A., Aldering, G., Amanullah, R., et al.\ 2003, \apj, 598, 102
\bibitem[Krueger et al.(2010)]{kru10} Krueger, B.~K., Jackson, A.~P., Townsley, D.~M., et al.\ 2010, \apjl, 719, L5
\bibitem[Kroupa(2001)]{kr01} Kroupa, P.\ 2001, \mnras, 322, 231
\bibitem[Lampeitl et al.(2010)]{lam10} Lampeitl, H., Smith, M., Nichol, R.~C., et al.\ 2010, \apj, 722, 566
\bibitem[Le Borgne \& Rocca-Volmerange(2002)]{zpeg} Le Borgne, D., \& Rocca-Volmerange, B.\ 2002, \aap, 386, 446 
\bibitem[Maiolino et al.(2008)]{mai08} Maiolino, R., Nagao, T., Grazian, A., et al.\ 2008, \aap, 488, 463
\bibitem[Mannucci et al.(2010)]{man10} Mannucci, F., Cresci, G., Maiolino, R., Marconi, A., \& Gnerucci, A.\ 2010, \mnras, 408, 2115 
\bibitem[Mannucci et al.(2011)]{man11} Mannucci, F., Salvaterra, R., \& Campisi, M.~A.\ 2011, \mnras, 414, 1263
\bibitem[Miknaitis et al.(2007)]{mik07} Miknaitis, G., Pignata, G., Rest, A., et al.\ 2007, \apj, 666, 674
\bibitem[Neyman \& Pearson(1933)]{nplemma}Neyman, J., \& Pearson, E. S., \ 1933, Phil. Trans. R. Soc. Lond. A 231, 289
\bibitem[Perez-Montero et al.(2012)]{per12} Perez-Montero, E., Contini, T., Lamareille, F., et al.\ 2012, arXiv:1210.0334 
\bibitem[Perlmutter et al.(1998)]{perl98} Perlmutter, S., Aldering, G., della Valle, M., et al.\ 1998, \nat, 391, 51
\bibitem[Perlmutter et al.(1999)]{perl99} Perlmutter, S., Aldering, G., Goldhaber, G., et al.\ 1999, \apj, 517, 565
\bibitem[Phillips(1993)]{phil93} Phillips, M.~M.\ 1993, \apjl, 413, L105
\bibitem[Riess et al.(1996)]{rie96} Riess, A.~G., Press, W.~H., \& Kirshner, R.~P.\ 1996, \apj, 473, 588
\bibitem[Riess et al.(1998)]{rie98} Riess, A.~G., Filippenko, A.~V., Challis, P., et al.\ 1998, \aj, 116, 1009 
\bibitem[Riess et al.(2004)]{rie04} Riess, A.~G., Strolger, L.-G., Tonry, J., et al.\ 2004, \apj, 607, 665
\bibitem[Riess et al.(2007)]{rie07} Riess, A.~G., Strolger, L.-G., Casertano, S., et al.\ 2007, \apj, 659, 98
\bibitem[Sako et al.(2008)]{sako08} Sako, M., Bassett, B., Becker, A., et al.\ 2008, \aj, 135, 348
\bibitem[Sako et al.(2011)]{sako11} Sako, M., Bassett, B., Connolly, B., et al.\ 2011, \apj, 738, 162
\bibitem[Salim et al.(2007)]{sal07} Salim, S., Rich, R.~M., Charlot, S., et al.\ 2007, \apjs, 173, 267 
\bibitem[Smith et al.(2011)]{smith10} Smith, M., Nichol, R.~C, Dilday, B., et al.\ 2011, arXiv:1108.4923
\bibitem[Sullivan et al.(2006)]{sul06} Sullivan, M., Le Borgne, D., Pritchet, C.~J., et al.\ 2006, \apj, 648, 868
\bibitem[Sullivan et al.(2010)]{sul10} Sullivan, M., Conley, A., Howell, D.~A., et al.\ 2010, \mnras, 406, 782
\bibitem[Sullivan et al.(2011)]{sul11} Sullivan, M., Guy, J., Conley, A., et al.\ 2011, \apj, 737, 102
\bibitem[Suzuki et al.(2012)]{suz12} Suzuki, N., Rubin, D., Lidman, C., et al.\ 2012, \apj, 746, 85
\bibitem[Timmes et al.(2003)]{tim03} Timmes, F.~X., Brown, E.~F., \& Truran, J.~W.\ 2003, \apjl, 590, L83
\bibitem[Tonry et al.(2003)]{ton03} Tonry, J.~L., Schmidt, B.~P., Barris, B., et al.\ 2003, \apj, 594, 1
\bibitem[Tremonti et al.(2004)]{tre04} Tremonti, C.~A., Heckman, T.~M., Kauffmann, G., et al.\ 2004, \apj, 613, 898
\bibitem[Wood-Vasey et al.(2007)]{wood07} Wood-Vasey, W.~M., Miknaitis, G., Stubbs, C.~W., et al.\ 2007, \apj, 666, 694
\bibitem[Yates et al.(2012)]{yates12} Yates, R.~M., Kauffmann, G., \& Guo, Q.\ 2012, \mnras, 422, 215
\bibitem[York et al.(2000)]{york00} York, D.~G., Adelman, J., Anderson, J.~E., Jr., et al.\ 2000, \aj, 120, 1579



\end{thebibliography}
\end{document}